\documentclass[preprint]{vldb}
\usepackage{graphicx}
\usepackage{balance}

\vldbTitle{Network-accelerated Distributed Machine Learning Using MLFabric}
\vldbAuthors{Raajay Viswanathan, and Aditya Akella}
\vldbDOI{}
\vldbVolume{}
\vldbNumber{}
\vldbYear{2020}

\usepackage{balance}
\usepackage{epsfig}
\usepackage{bm}
\usepackage{times}
\usepackage{datetime}
\usepackage{listings}
\usepackage{xcolor}
\usepackage[normalem]{ulem}
\usepackage{microtype}
\usepackage{multirow,makecell}
\usepackage{url}
\def\outlineenabled{Compile with outline text}
\def\commentenabled{Include comments in the draft}

\usepackage{grffile}

\usepackage{graphics,graphicx}

\usepackage{amsmath,amssymb}

\usepackage{color, colortbl}

\usepackage{xparse}

\usepackage[]{cleveref}

\usepackage[
  labelfont=footnotesize,
  textfont=footnotesize,
  justification=justified,
  skip=0.5\baselineskip plus 1\baselineskip minus 1\baselineskip
]{subcaption}

\usepackage[
  labelfont={bf,small},
  textfont=small,
  justification=justified,
  skip=0.5\baselineskip plus 1\baselineskip minus 1\baselineskip,
]{caption}

\usepackage{xspace}

\usepackage{boxedminipage2e}

\usepackage[ruled, linesnumbered, noline]{algorithm2e}

\usepackage{epstopdf}

\usepackage{tikz}
\usetikzlibrary{backgrounds}
\usetikzlibrary{calc}
\usetikzlibrary{positioning}
\usetikzlibrary{automata}
\usetikzlibrary{shapes}
\usetikzlibrary{arrows}
\usetikzlibrary{shadows}
\usetikzlibrary{decorations.pathreplacing}
\usetikzlibrary{decorations.text}
\usetikzlibrary{shapes.multipart}
\usetikzlibrary{fit}
\usetikzlibrary{patterns}

\usepackage{pgfplots}
\usepgfplotslibrary{units}

\makeatletter
\tikzset{nomorepostaction/.code=\let\tikz@postactions\pfgutil@empty}
\makeatother

\input{./arxiv-latexdefs/mathdefs}
\input{./arxiv-latexdefs/envdefs}
\input{./arxiv-latexdefs/namedefs}

\pgfplotsset{compat=1.14}
\newcommand{\sysname}{{ML}{f}abric\xspace}

\newcommand{\name}{{ML}{f}abric\xspace}
\newcommand{\coordinator}{{scheduler}\xspace}
\newcommand{\Coordinator}{{Scheduler}\xspace}

\SetNoFillComment
\SetCommentSty{mycommfont}

\SetAlCapSty{xAlCapSty}

\makeatletter
\newcommand{\removelatexerror}{\let\@latex@error\@gobble}
\makeatother

\begin{document}

%don't want date printed
\date{}

%make title bold and 14 pt font (Latex default is non-bold, 16 pt)
\title{\Large \bf Network-accelerated Distributed Machine Learning Using \sysname}

\numberofauthors{2}
\author{
\alignauthor
Raajay Viswanathan\\
\affaddr{University of Wisconsin Madison}\\
\email{raajay@cs.wisc.edu}
\alignauthor
Aditya Akella\\
\affaddr{University of Wisconsin Madison}\\
\email{akella@cs.wisc.edu}
}

\maketitle
% !TEX root = main.tex
\noindent
{\bf Abstract:} Existing distributed machine learning (DML) systems
focus on improving the computational efficiency of distributed learning,
whereas communication aspects have received less attention. Many DML systems treat the network as a blackbox. Thus, DML
algorithms' performance is impeded by network bottlenecks, and DML systems
end up sacrificing important algorithmic and system-level
benefits.  We present \sysname{}, a communication library that manages
all network transfers in a DML system, and holistically determines the
communication pattern of a DML algorithm at any point in time. This
allows \sysname{} to carefully order transfers (i.e., gradient
updates) to improve convergence, opportunistically aggregate updates
in-network to improve efficiency, and proactively replicate some of
them to support  new notions of fault tolerance. We
empirically find that \sysname{} achieves up to $3\times$ speed- up
in training large deep learning models in realistic dynamic cluster
settings. 

% !TEX root = main.tex
\section{Introduction}
\label{sec:introduction}

Machine learning (ML) is revolutionizing not only the computing
industry, but also fields such as healthcare and education, where
ML techniques are driving key
applications.  Thus, there is a race to build new ML
systems~\cite{tensorflow-osdi,caffe2-online,pytorch,bosen} that
efficiently learn complex models from big datasets.

To support large model sizes and training data most systems execute
distributed versions of ML (DML) algorithms across 10s-100s of
workers in a cluster.  These DML algorithms, e.g., distributed
stochastic gradient descent (SGD), and distributed Latent Dirichlet
Allocation (LDA)~\cite{blei-lda,plda}, are iterative in nature, and
are both computation and communication intensive.  In each iteration,
a worker computes an update to the large model, which then needs to be
disseminated to all other workers.  Model updates can be $\ge100$MB
per worker per iteration, yielding large network transfers
(\secref{sec:background}).

Many DML systems~\cite{tensorflow-osdi,pytorch,geeps,caffe2-online}
focus on addressing the performance of computing updates at individual
workers, e.g., via optimal use of hardware
accelerators~\cite{tiresias,optimus}.  In contrast,
systematically addressing communication efficiency and network
bottlenecks has received limited attention.
In most systems, the application (DML algorithm) manages both
computation and communication. A simple network view, as offering
fixed bandwidth between all cluster workers, is
adopted. Application-level techniques are then used to reduce total
data transferred to/from a worker to avoid network bottlenecks.

DML systems thus treat the network as a blackbox, and as such, are
unable to overcome network issues.  Consider Parameter
Server (PS) based ML systems~\cite{bosen,mxnet-arxiv}. The
model is stored at a separate location (server); in every iteration
workers pull the latest model and compute an update, which is then
shipped to the server and applied to the model. PS systems support
flexible consistency schemes, e.g., strict synchronous, stale
synchronous, or asynchronous model updates (\xref{sec:background}),
which help improve DML algorithms' compute efficiency and
convergence~\cite{parameterserver-osdi}.  However, they deal with
network efficiency using \textit{ad hoc} application-level approaches,
such as, dropping updates deemed not significant, or coarsely
quantizing updates~\cite{bosen}. Unfortunately, these approaches
affect algorithm convergence~\cite{one-bit-sgd}, and yet may not be
effective enough in dealing with serious onset of congestion.

Likewise, MPI-based systems~\cite{caffe2-online,distributed-pytorch}
--- which support only synchronous SGD
  --- employ \textit{MPI\_AllReduce} operations to minimize data
  communication. Updates are aggregated along a static topology
  (e.g., a ring or a tree) among the workers. Unfortunately,
  network-unawareness of the aggregation topology means that a worker
  stuck behind a network bottleneck will block the aggregation of
  updates from other workers.

Treating the network as a blackbox imposes other drawbacks. In
particular, DML systems leave on the table potential algorithmic
improvements and new framework-level support for further improving
large-scale DML that can be achieved by {\em actively leveraging} the
network. We argue these issues in the context of \sysname{}, a new DML
communication library we have built. \sysname{} applies equally to PS
and MPI systems.  \sysname{} decouples computation from
communication. Applications hand over the entire responsibility of
transferring the model and its updates across the network to
\sysname{}. \sysname{} then holistically determines the communication
pattern of a DML algorithm at any given time in a network-aware
fashion. This offers three benefits.

\noindent
{\bf 1. Flexible aggregation to overcome network bottlenecks.} Using
holistic control, \sysname{} can determine {\em in-network}
aggregation strategies. Workers can be dynamically organized into
tree-like topologies over which updates are routed and aggregated
before being committed at a server. This helps improve network
efficiency in the presence of dynamically changing compute or network
contention, which is common in shared
environments~\cite{tiresias,gandiva}. It is orthogonal to the
algorithm-level approaches above (e.g., update quantization).

\noindent
{\bf 2. Leveraging the network for algorithmic advances} In
asynchronous SGD, updates from slow workers,
e.g., compute stragglers, or those stuck behind a network bottleneck,
have a high \textit{delay}, i.e., their update is computed from an old
model version.  Applying stale updates to the model can affect
convergence~\cite{delayedsgd}.  To address this, asynchronous
algorithms set small learning rates based on the worst case delay
observed, which slows down training. By leveraging control over
communication, \sysname{} can orchestrate how and when updates are
transferred over the network, thereby explicitly {\em controlling} the
staleness of each update, and bounding worst case delay
distribution. This allows the application to set a high learning rate
even under a changing execution environment, which improves
convergence. Further, updates with high delay that negatively affect
convergence can be dropped at the worker itself without wasting
network resources.

We find that using \sysname{}'s in-network aggregation and explicit
control over update delay have other surprising algorithmic
consequences. Empirically, for some popular large deep neural net
models (e.g., ResNet-50), we find that these techniques help
asynchronous SGD-based training atop PS frameworks to converge {\em faster}
than synchronous SGD-based training atop MPI in some straggler
settings. The latter training approach has been the de facto standard
for deep learning because of the significant network bottlenecks faced
by the naive use of asynchronous algorithms and/or PS; many
foundational systems and algorithmic approaches have optimized the
speed of deep learning in MPI settings. In contrast, the use of
\sysname{} makes asynchronous/PS now a contender, and opens the door
to related new systems and algorithmic approaches to improving deep
learning.

\noindent
{\bf 3. Leveraging the network for framework improvements} Existing PS systems~\cite{parameterserver-osdi}
use a hot-standby for server fault tolerance.  Chain replication is
employed to ensure every model update to the parameter server is also
applied to the replica, enforcing strong consistency. However, chain
replication introduces additional per-iteration latency, and
exacerbates network contention at the server. \sysname{}'s control
over communication supports flexible \emph{bounded consistency}, which
helps significantly control replication overhead. Under bounded
consistency, we require that $||\wb_{s} - \wb_{r}|| \leq {Div}_{max}$
should always be satisfied, where $\wb_{s}$ and $\wb_{r}$ denote the
models stored at the server and replica, and $Div_{max}$ is a
user-configured divergence limit; higher $Div_{max}$ leads to lower
replication overhead, but higher recovery cost. Bounded consistency is
sufficient for ML algorithms due to their stochastic nature; upon
server failure, the lost work due to non-replicated updates can be
recovered by generating fresh worker updates using the latest model at
the replica.  In \sysname{},
workers replicate updates, and the network controls them to carefully
schedule original and replica transfers so as to ensure that
divergence stays within $\leq {Div}_{max}$.

We implement \sysname{} as a thin communication layer between DML
applications~\cite{keras,tensorflow-osdi} and MPI communication
libraries~\cite{openmpi,nccl}.
It internally uses MPI APIs to
aggregate/schedule transfers across the network and/or to a server.

In designing \sysname{}, we make four technical contributions. First,
we {\em prove} that the convergence of asynchronous SGD for convex
loss functions improves when we actively bound the variance of the
delay distribution (\secref{sec:motivation}). Our evaluation shows
empirically that bounding the delay speeds up convergence even for non-convex optimization problems
used in training deep neural networks and other asynchronous DML
algorithms like distributed LDA using Gibbs sampling
(\xref{sec:evaluation}).  Second, we design a {\em scheduling
  algorithm} that, given a set of worker updates, computes the update
transfer schedules in a network-aware fashion
(\secref{sec:method}). This algorithm transfers updates at
non-overlapping times, reserving bandwidth per transfer, and it
carefully orders worker updates. We show that the former enables  a
fast rate of model updates, and ordering helps bound delays.
Third, we develop
an in-network {\em aggregation algorithm} that determines whether to
send each update directly to a server, or to an aggregator first. It
performs the best in-network aggregation possible while efficiently
utilizing aggregators' bandwidth and preserving update
ordering. Finally, we develop a {\em replication algorithm} that
opportunistically schedules transfers to a replica server while
leveraging spare capacity. It prioritizes primary server transfer
schedules and while always bounding divergence.
\sysname{}'s scheduler runs these three
algorithms in sequence for every batch of updates, leading to
delay-bounded, divergence-bounded, networking efficient fast model
updates.

We evaluate \sysname{} using a 30 worker cluster with P100 GPUs and
quad-core CPUs under 9 different time-varying network and compute
straggler settings (\secref{sec:evaluation}). We study large deep
neural net (ResNet-50 and ResNet-152) training, and distributed LDA
for topic modeling using Gibbs sampling.  We show that \sysname{}
improves overall training times by $1.2-3\times$ compared to
state-of-the-art under various straggler settings.
\sysname{} offers up to 30X lower replication overhead for PS systems
in some scenarios.

% !TEX root = main.tex
\providecommand{\tconvsyn}{}
\renewcommand{\tconvsyn}{t_{syn}}
\providecommand{\nconvsyn}{}
\renewcommand{\nconvsyn}{N_{syn}}

\providecommand{\tconvasyn}{}
\renewcommand{\tconvasyn}{t_{asyn}}
\providecommand{\nconvasyn}{}
\renewcommand{\nconvasyn}{N_{asyn}}

\providecommand{\syncsgd}{}
\renewcommand{\syncsgd}{\textit{SyncSGD}\xspace}
\providecommand{\asyncsgd}{}
\renewcommand{\asyncsgd}{\textit{AsyncSGD}\xspace}

\section{DML Performance Analysis}
\label{sec:background}
\label{sec:analyze_runtimes}
\label{sec:performance-analysis}

The \emph{de facto} algorithm of choice for various ML applications
like Deep learning, Generalized Linear Models, etc., is Stochastic
Gradient Descent (SGD)~\cite{nocedal-wright}.  SGD is inherently
serial; in each iteration the model is updated using a gradient from a
single sample or a \emph{mini-batch} of training
data~\cite{bottou-lsml}. In order to distribute SGD, ML practitioners
have successfully used its different variants, each having different
model consistency requirements: (\#1) asynchronous
SGD~\cite{hogwild,delayedsgd}, (\#2) stale synchronous SGD~\cite{ssp},
and (\#3) synchronous SGD~\cite{nocedal-wright}.

Our primary focus in this paper is on \#1 as realized in parameter
server (PS) DML systems.  The entire suite of \sysname{}'s algorithms
for network control (scheduling; in-network aggregation;
bounded-divergence replication) apply to \#1.
Subsets of \sysname{} also apply to
\#2 (both PS and MPI) and \#3; we discuss these
in \secref{sec:other-aspects}, and evaluate
in \secref{sec:evaluation}. Furthermore, in \secref{sec:evaluation},
we show \sysname{}'s benefits for other (non-SGD)
synchronous/asynchronous algorithms like distributed LDA.

We provide a brief overview of \#1 below, followed by DML algorithm's
computation and communication characteristics.

\noindent
\textbf{Asynchronous SGD:} Here, a worker computes a gradient update using a mini-batch of
local data, pushes it to the server and pulls the latest model.  The update from each worker is applied independently
to the model at the server at each iteration.  In iteration $t$, the
update computed by a worker and the model update at the server,
respectively, are:
% \begin{small}
\begin{eqnarray}
  u^j_{t} &=& - \eta \frac{\partial}{\partial \wb} L(D^j, \wb_{t-\tau}) + \lambda (\wb_{t - \tau})
  \label{eq:async-sgd-worker-compute} \\
  \wb_{t+1} &=& \wb_t + u^j_t + \gamma \{\wb_{t} - \wb_{t-1}\}
  \label{eq:async-sgd-server-compute}
\end{eqnarray}
% \end{small}
where, $\wb_t$ is the model after iteration $t$, $L$ is the loss function,
$D^j$ is the mini-batch at worker $j$, $\lambda(\wb)$ is the regularization
term, $\eta$ is the learning rate and $\gamma$ is the momentum
term\footnote{This update strategy corresponds to SGD with momentum which is
shown to be beneficial for asynchronous SGD~\cite{yellowfin}}.

The update, $u_t^j$, is calculated using an old version of the model,
$\wb_{t-\tau}$ instead of $\wb_t$. Here, $\tau$ is called the {\bf
delay} of update $u_t^j$; it is the difference between the version of
the model being updated and the one used to compute the update.

\noindent
\textbf{Performance based on model complexity:}
Most real-world models trained with DML algorithms are complex.  Consider the
image recognition neural network model, ResNet50, that
achieves up to 75\% accuracy in classifying images among 1000 classes. It is
100MB in size; however, GPUs (e.g., NVIDIA P100) can process up to 200 images
per second to compute updates for the model. In a distributed training setup
(see~\secref{sec:evaluation}), the number of images used to compute an update
at a worker is much lower; typically 32 images. In such a scenario, we
find that the computation phase at a DML workers takes less than 100ms. On the
other hand, faster compute means that workers have to exchange 100MB worth of
updates amongst themselves every 100ms. This causes high communication
overhead; even bandwidth optimal AllReduce strategies like ring AllReduce (used
in synchronous SGD algorithms) take at least 320ms (3$\times$ computation
time) to exchange all updates when all workers are connected by commodity
10Gbps Ethernet. By aggregating updates between GPUs in a worker before exchanging over
the network the communication cost is reduced to 160ms; further, effective
pipelining of computation and communication reduces the overall time per
iteration from (100 + 160)ms to 200ms.  Other update exchange strategies
(e.g., binary tree AllReduce or Parameter Server with single server) can have at least $20\times$
communication overhead. Increasing the number of parameter servers reduces the
network load per server (even though it is still higher than ring AllReduce), at the cost of increased communication between servers
to propagate model version information. This makes it undesirable for
asynchronous ML algorithms.

Similar trend is observed for other algorithms; e.g., distributed
LDA.  Topic modeling using LDA on the NY Times
dataset~\cite{nytimes} with 32 parallel workers (see~\secref{sec:evaluation})
has computation cost of 180ms at each worker. The communication cost
(time to exchange updates) assuming ring AllReduce is 160ms. However,
for PS-based system with 10Gbps server bandwidth, communication cost
is $\sim$1.8s, i.e., $10\times$ computation.

\noindent \textbf{Performance with stragglers and network bottlenecks:}
For synchronous algorithms, where the progress is determined by the slowest
worker, the effect of stragglers and network bottlenecks is prominent. We find
(\secref{sec:evaluation}) that the per-iteration time increases $8\times$ when
10\% of the workers took 4X longer in each iteration and 10\% of the incoming
and outgoing network links had bandwidth lowered to $5$Gbps.

In asynchronous SGD, these slow workers observe high delays. This
severely impacts convergence speed or converged model accuracy (\secref{sec:evaluation}). We find that
asynchronous LDA, upon introducing network bandwidth fluctuations
(bandwidth on 10\% of the links are reduced to 5 Gbps every 5s), takes
35\% more iterations to converge due to network stragglers.

% !TEX root = main.tex
\section{Central ideas}
\label{sec:motivation}

Today's DML systems' network-agnosticity causes slowdowns in the face of
compute or network contention (stragglers).
In \sysname{}, instead of treating the network as a blackbox,
\emph{all} transfers of a DML algorithm are handed off to a
communication library, which determines the entire communication
pattern at any point in time. For simplicity, we
explain \sysname{} in the context of PS systems and asynchronous
algorithms.

In \sysname{}, each update \emph{push} from a worker to a server is intercepted and
fulfilled later, at which time it is either directly forwarded to the
server or passed through intermediate hops where the update is
aggregated with other workers' updates.  We refer to the ability to
finely control the transfer of model and its update over the network
as \emph{in-network control}. In this section, we explain the benefits
of in-network control using theory and qualitative
arguments. Algorithms that realize the benefits are presented in
\secref{sec:method}.

First, in-network control enables \emph{network-based delay
  management} (\secref{sec:delay-management-motivation}) -- i.e.,
managing delay observed at the server by controlling the order of
updates inside the network. This helps asynchronous SGD by
lowering the number of iterations to convergence as well as the
average iteration time.  Second, network control enables in-network
aggregation of updates (\secref{sec:aggregation-motivation}), which
further improves per-iteration performance.  Third, it enables
off-loading model replication from parameter servers to the
network by ensuring consistent ordering of updates across primary and
replica servers, and bounding model-replica divergence
(\secref{sec:in-network-replication}). This relieves both server-side
and network replication load while enabling fast recovery from server
failure.

\subsection{Delay management}
\label{sec:theorem}
\label{sec:delay-management-motivation}

We describe what delay management is and how it is helpful. Then, we
make a case for it to be {\em network-based}.

Recall from eqns. \ref{eq:async-sgd-server-compute}
and~\ref{eq:async-sgd-worker-compute}, that workers' updates are applied to
the model in a delayed fashion.
\cite{delayedsgd}
shows that for well behaved convex loss functions asynchronous SGD converges as long as the
delay for each update is bounded ($\tau < \tau_{max}$)
and the learning rate or ``step size'' in iteration, $t$, is set as: $\eta =
\frac{C}{\sqrt{\tau_{max} t}}$, where $C$ is a constant.
As a result, for execution environments with large observed delays the
learning rate must be set small to guarantee convergence. This increases the number of iterations until convergence.
In response, \cite{adadelay} advocates on making learning
rate a function of the delay observed for a worker; under the
assumption that the delay follows an uniform distribution, $\tau \in
\text{Uniform}[0, 2 \bar{\tau}]$, they show that delay adaptive
SGD converges as:
% \begin{small}
\begin{equation}
  E[L(\wb_t)] - L(\wb^*) \leq \mathcal{O} \left(\frac{\bar{\tau} \sqrt{t}}{t}\right)
\end{equation}
% \end{small}
where, $w^*$ is the optimal model minimizing loss function $L(.)$, and
$\hat{\wb}(t)$ is the estimated model after $t$ iterations.

Building on this result, we show that (\xref{sec:proof-convergence}) if $\tau \in
\text{Uniform}[\bar{\tau} - \epsilon, \bar{\tau} + \epsilon]$, delay
adaptive SGD converges as:
% \begin{small}
\begin{equation}
  E[L(\wb_t)] - L(\wb^*) \leq \mathcal{O} \left(\frac{\epsilon \sqrt{t + \bar{\tau} - \epsilon}}{t}\right)
  \label{eq:convergence-mlfabric}
\end{equation}
% \end{small}
In other words, we can get constant factor speedup in convergence,
where the speedup is inversely proportional to $\epsilon$.

Based on this result, our first idea is to {\em carefully control the
  order} in which updates are applied to the model. This reduces the
variance of the delay distribution in asynchronous SGD and bounds
maximum delay.

\begin{figure*}[!t]%{{{
  \begin{tabular}{ccc}
    \begin{boxedminipage}[t]{0.31\textwidth}%
      \centering%%
      \includegraphics[width=0.7\linewidth]{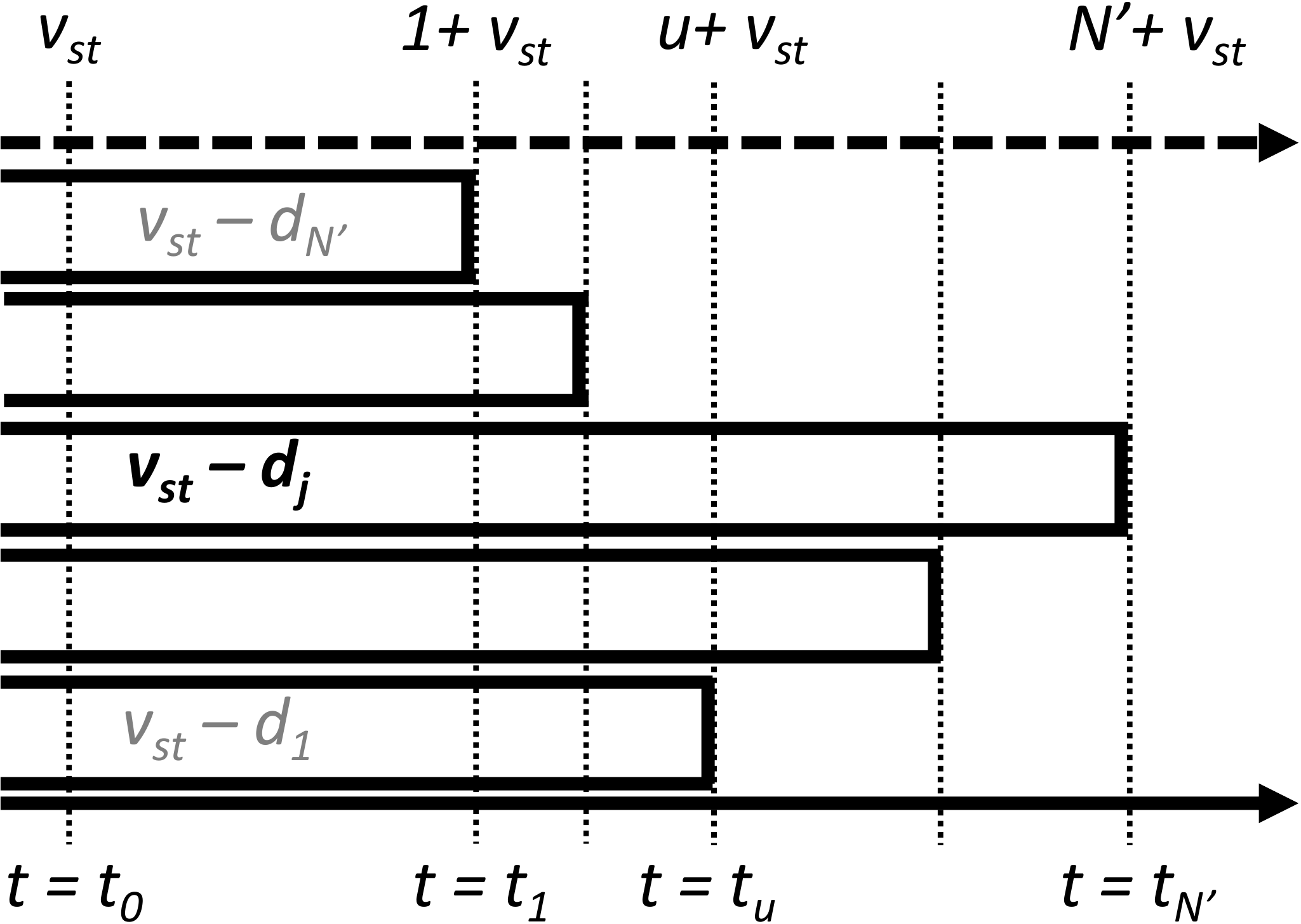}%%
      \subcaption{Delay bound is violated for one update%%
      \label{fig:update-sharing}}%%
    \end{boxedminipage}%
    &%%
    \begin{boxedminipage}[t]{0.31\textwidth}%
      \centering%%
      \includegraphics[width=0.7\linewidth]{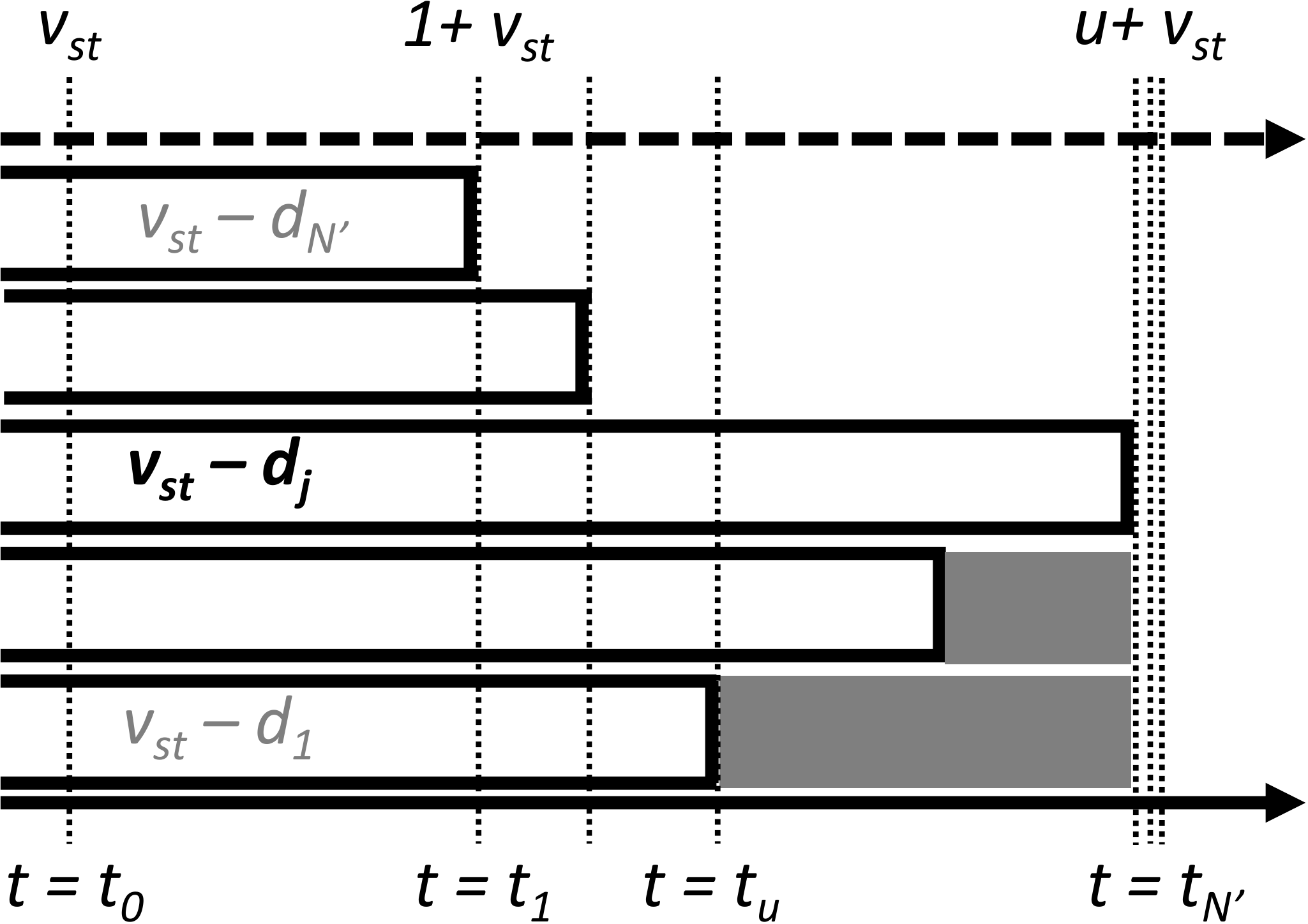}%%
      \subcaption{Buffering updates to bound delay at server%%
      \label{fig:update-buffering}}%%
    \end{boxedminipage}%
    &%%
    \begin{boxedminipage}[t]{0.31\textwidth}%
      \centering%%
      \includegraphics[width=0.7\linewidth]{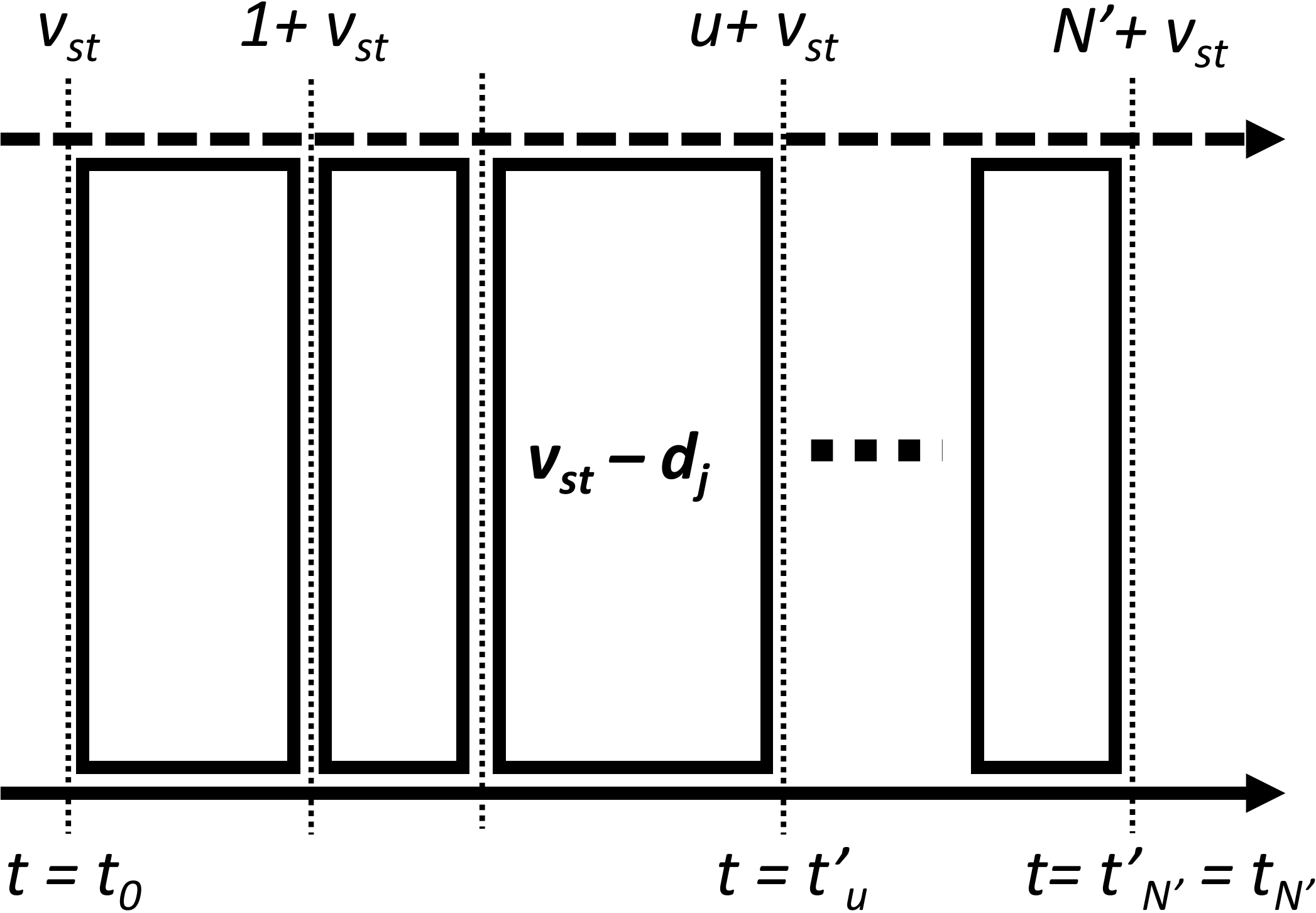}%%
      \subcaption{Ordering gradient updates over the network%%
      \label{fig:update-ordering}}%%
    \end{boxedminipage}%
  \end{tabular}%%
  \caption{Timeline of gradient transfers and model updates for different
  scenarios. In (a) we show the situation today where all $N'$ workers transfer
their updates concurrently over the network. Let us assume that
network bandwidth is shared, and that the server updates the model
using updates in the order in which their network transfer
completes. Figure (a) shows the time line for one
such scenario; note that update $j$ which is computed with the oldest
version of the model completes last. As a result, the observed delay
at the server for update $j$ is: $v_{st} + N' - (v_{st} - \tau_{max} +
u) = \tau_{max} + (N' - u)$. Since, $u < N'$, the delay $ >
\tau_{max}$. \label{fig:update-timeline}}
  % \vspace{-0.1in}
\end{figure*}

\subsubsection{In-network Control: Fresher Model Versions}
\label{sec:time-sharing}
\label{sec:re-ordering}

The above ordering of updates can be realized at the parameter
server. However, we argue that it is better to leverage in-network
control, thereby enforcing {\em network-based} ordering. This is because
in-network control helps make fresher model versions available
earlier, as argued below.

Consider an execution of asynchronous SGD; let us assume that at some
time instant, $t=t_0$, $N'$ ($< N$ total) workers have pending
gradient updates that need to be applied to the model. These updates
were computed using prior versions of the model, versions $[v_{st} -
  d_1, \ldots, v_{st} - d_{N'}]$, where $v_{st}$ is the current
version of the model and $d_{i}, 1 \leq i \leq N'$, are integers
denoting the delay of the update if it is applied to the model at
version $v_{st}$. Assume that there exists exactly one $j$, such that
$d_{j} = \tau_{max} - u, u < N'$ and $d_{i} < \tau_{max} - N', \forall
i \neq j$, where $\tau_{max}$ is the maximum allowable delay. In other
words, one of the updates ($j$) has been computed with an older version of
the model when compared to others.

In Figure~\ref{fig:update-sharing}, we show how updates are
transferred today, and how that may cause delay to exceed $ \tau_{max}$
(see caption).

As one alternative, we can enforce that no update with
delay $>\tau_{max}$ should be applied to the model; this causes update
$j$ to be discarded, resulting in lost work.

Another alternative is
server-based update ordering, where we buffer updates that complete
after $t_{u}$  at the server (fig.~\ref{fig:update-buffering}), and
apply them after update $j$ has been transferred and applied to the
model. This ensures that update $j$'s delay is exactly $\tau_{max}$;
the delay for all buffered updates increases by $1$ but remains under
$\tau_{max}$. The downside is that the workers' interim pull requests do not see
new model versions: all pull requests for the model between
$[t_{u},t_{N'})$ are returned version $v_{st} + u$, which is worse
  than in fig.~\ref{fig:update-sharing}.

The final alternative is {\em in-network control}, where we can
enforce {\em network time sharing}, i.e., different updates are
transmitted by the network at carefully-chosen non-overlapping times
at bottleneck links (See fig.~\ref{fig:update-ordering}; note: we assume a
single bottleneck at the server here). The total time to transfer all
the updates would be the same ($t'_{N'}= t_{N'}$) since the total data
transferred over the network is same as before.  However, as long as
update transfers are scheduled such that update $j$ is transferred as
one of the first $u$, the delay bound will be satisfied without any
need for server-side buffering. Further, by transferring updates $1$
through $N'$ except update $j$ in the {\em order of completion time},
we can emulate shorted-job-first and minimize average update
time. This makes new model versions available earlier than in
figs.~\ref{fig:update-sharing} and \ref{fig:update-buffering}. Our
update scheduling algorithm in \secref{sec:method-ordering-updates}
relies on this idea.

\subsection{Aggregation}
\label{sec:aggr}
\label{sec:aggregation-motivation}

\begin{figure}%%
  \centering%%
  \begin{tabular}{@{}c@{}c@{}}%%
    \centering%%
    \begin{boxedminipage}[b]{0.49\linewidth}%%%
      \centering%%
      \includegraphics[width=\linewidth]{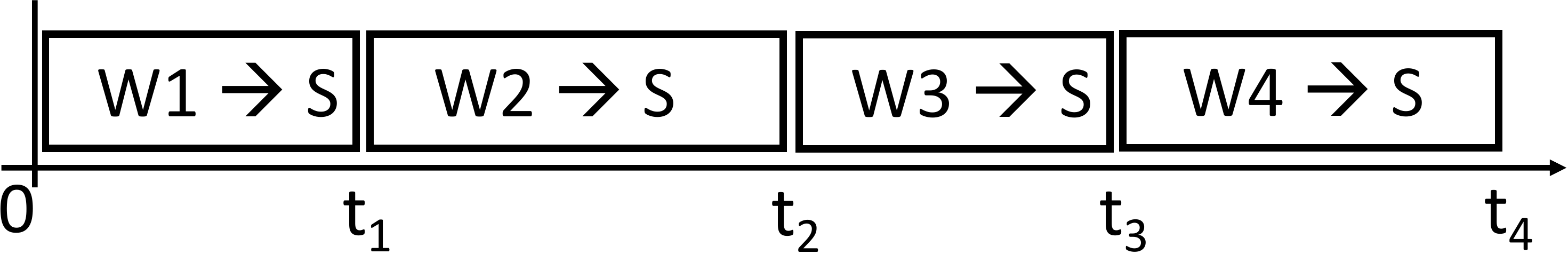}%%
      \subcaption{Updates forwarded directly to server \label{fig:no-aggregation}}%%
    \end{boxedminipage}%%%%
    &%%%%
    \hspace{0.02\linewidth}%%%%
    \begin{boxedminipage}[b]{0.49\linewidth}%%%
      \centering%%
      \includegraphics[width=\linewidth]{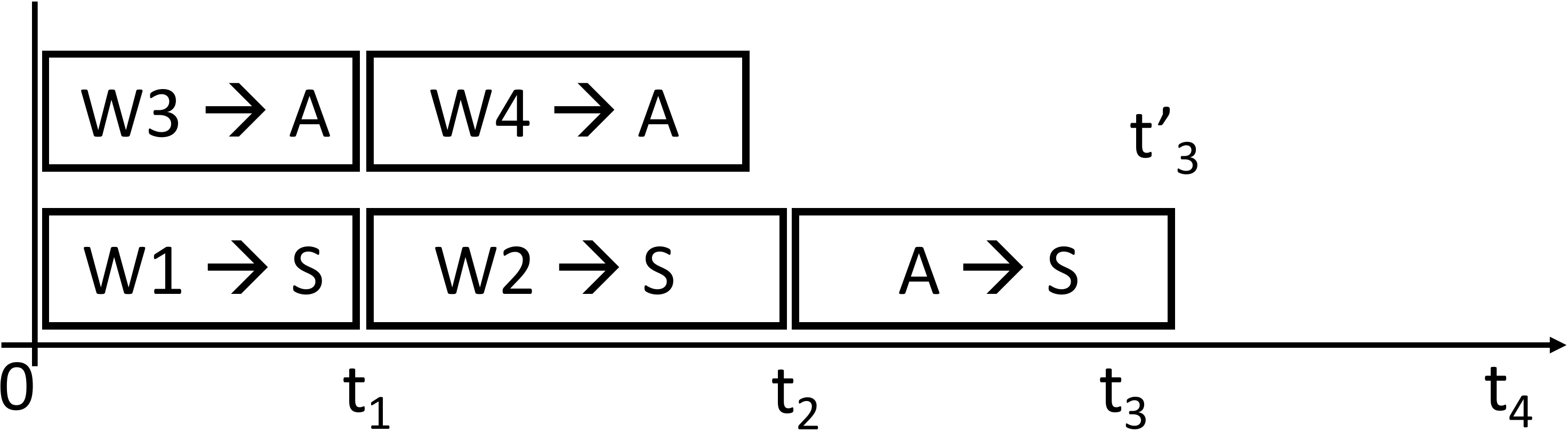}%%
      \subcaption{Update forwarded to server through aggregator \label{fig:aggregation}}%%
    \end{boxedminipage}%%%
  \end{tabular}%%
  \tightcaption{Example highlighting advantages of gradient aggregation}%%
  % \vspace{-0.1in}
\end{figure}%%

%% \noindent
%% {\bf 2. Even fresher updates via aggregation:}
In-network control enables the above ordered
updates that are ready to be sent to the server  to be further {\em
  opportunistically aggregated} at network locations before being
applied at the server. Thus, network load is lowered, and  model updates occur faster (at earlier times) w.r.t. not
aggregating in-network.

Say at time $t=0$ updates ($g_1, .., g_4$) from 4 different workers
are available. If all the updates are transferred to the server
directly in a time shared manner, completion times are as shown in
fig.~\ref{fig:no-aggregation}.  Even though the update from worker
$w_3$ is available at $t_0$, it is queued and starts transmitting only
at $t_2$ (after update from $w_2$ has completed).

With in-network control, say updates $g_1$ and $g_2$ are forwarded
directly to the server, but $g_3$ and $g_4$ are aggregated at $A$
(fig.~\ref{fig:aggregation}). Our aggregation algorithm in
\secref{sec:in-network-aggregation} constructs such aggregation topologies dynamically
based on current network load. Assuming full-bisection bandwidth,
$g_3$ and $g_4$ will be transferred concurrently with $g_1$ and $g_2$.
After aggregation, the result $r$ can be transferred to the server,
where it updates the model at $t'_3 < t_4$.

Since $|r| < |g_3| + |g_4|$, server network load is reduced. Also,
pull requests at time $t \geq t'_3$ can be replied with fresh
information of all 4 model updates; without aggregation, pull requests
within $[t'_3,t_4]$ don't capture the last update.

\subsection{Toward Bounded Consistency Replication}
\label{sec:in-network-replication}

In PS-based systems, the server stores the entire model. It is
therefore crucial to ensure server fault-tolerance.\footnote{Server
  fault-tolerance is needed because individual workers often do not
  read the entire copy of the model in each iteration (e.g., sparse
  logistic regression~\cite{parameterserver-osdi}), and/or because,
  apart from the model, parameter servers also store additional states
  not visible to workers like history of updates, prior learning rate,
  etc. used for momentum based model updates.}
Existing PS implementations use chain replication for fault
tolerance~\cite{chain-replication}, which incurs $O(k)$ data overhead
for $k$ replicas.
They attempt to reduce the data overhead by aggregating updates at the
server and forwarding them to the replica once every $n$
iterations. However, if updates are sparse, infrequent replication
only amortizes the server data overhead since the total data
transferred from server to replica is not reduced.

To reduce server load, we can enable a replication strategy based on workers
forwarding a copy of each update directly to the replica. However,
such worker-based replication is not easy to achieve without active
in-network control. In particular, having workers replicate by
themselves fundamentally cannot preserve the ordering of
updates. Coupled with the stateful nature of model updates
(eqn.~\ref{eq:async-sgd-server-compute}), this can result in unbounded
server-replica model divergence, which makes recovery from the replica
slow, if not impossible. We show this next. Then, we discuss how
in-network control helps.

Assume at time $t=0$, the server and
the replica contain identical models (i.e., $\wb^s_0 = \wb^r_0 =
\wb_0$) and have the same prior update ($h^0$).  Let $u_1, u_2$ be the
next two updates to the model at the server; assume the same updates
are applied in a different order ($u_2, u_1$) at the replica.  Then,
by applying eqn.~\ref{eq:async-sgd-server-compute} twice, the model at
server and replica can be computed as:
% \begin{small}
\begin{eqnarray}
  \wb^s_2 &=& \wb_0 + (\gamma h^0 + u_1) + \{\gamma (\gamma h^0 +  u_1) + u_2\} \nonumber \\
  \wb^r_2 &=& \wb_0 + (\gamma h^0 + u_2) + \{\gamma (\gamma h^0 +  u_2) + u_1\}
\end{eqnarray}
% \end{small}
Thus, the divergence is:
% \begin{small}
\begin{eqnarray}
  \wb^s_2 - \wb^r_2 &=& \gamma (u_1 - u_2) \nonumber \nonumber \\
  ||\wb^s_2 - \wb^r_2||_2 &=& \gamma  ||u_1 - u_2||_2
\end{eqnarray}
% \end{small}
Each such re-ordering of updates will add a further non-zero divergence between the server and replica!

\begin{figure}[t]%{{{
  \centering
  \begin{tabular}{@{}c@{}c@{}}
    \begin{boxedminipage}[t]{0.5\linewidth}
      \includegraphics[width=0.75\textwidth]{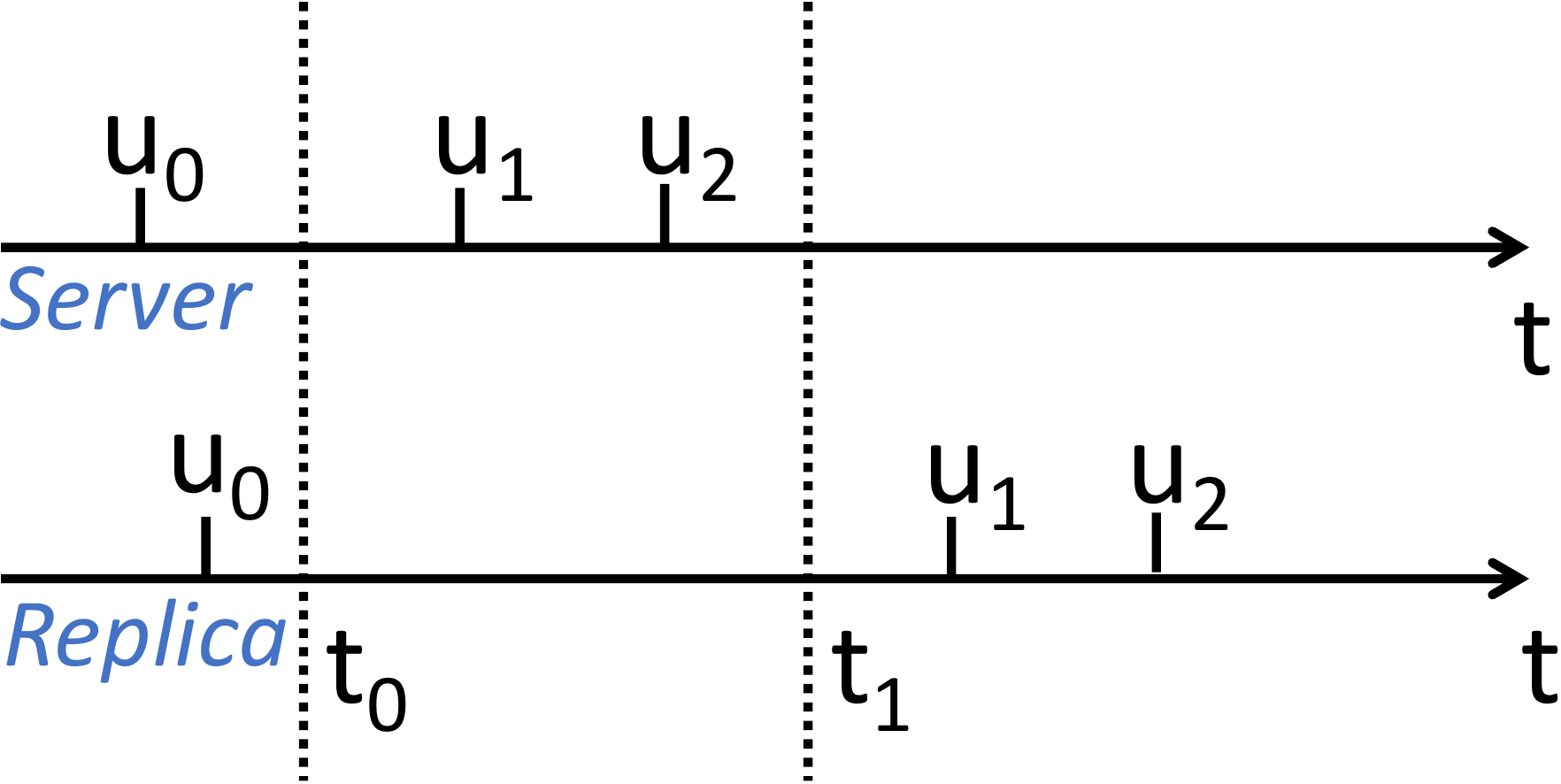}
      \subcaption{
        Replica lags the server by two updates
        \label{fig:update-schedule-a}
      }
    \end{boxedminipage}%%
    &%%
    \hspace{0.02\linewidth}%%
    \begin{boxedminipage}[t]{0.5\linewidth}
      \includegraphics[width=0.75\textwidth]{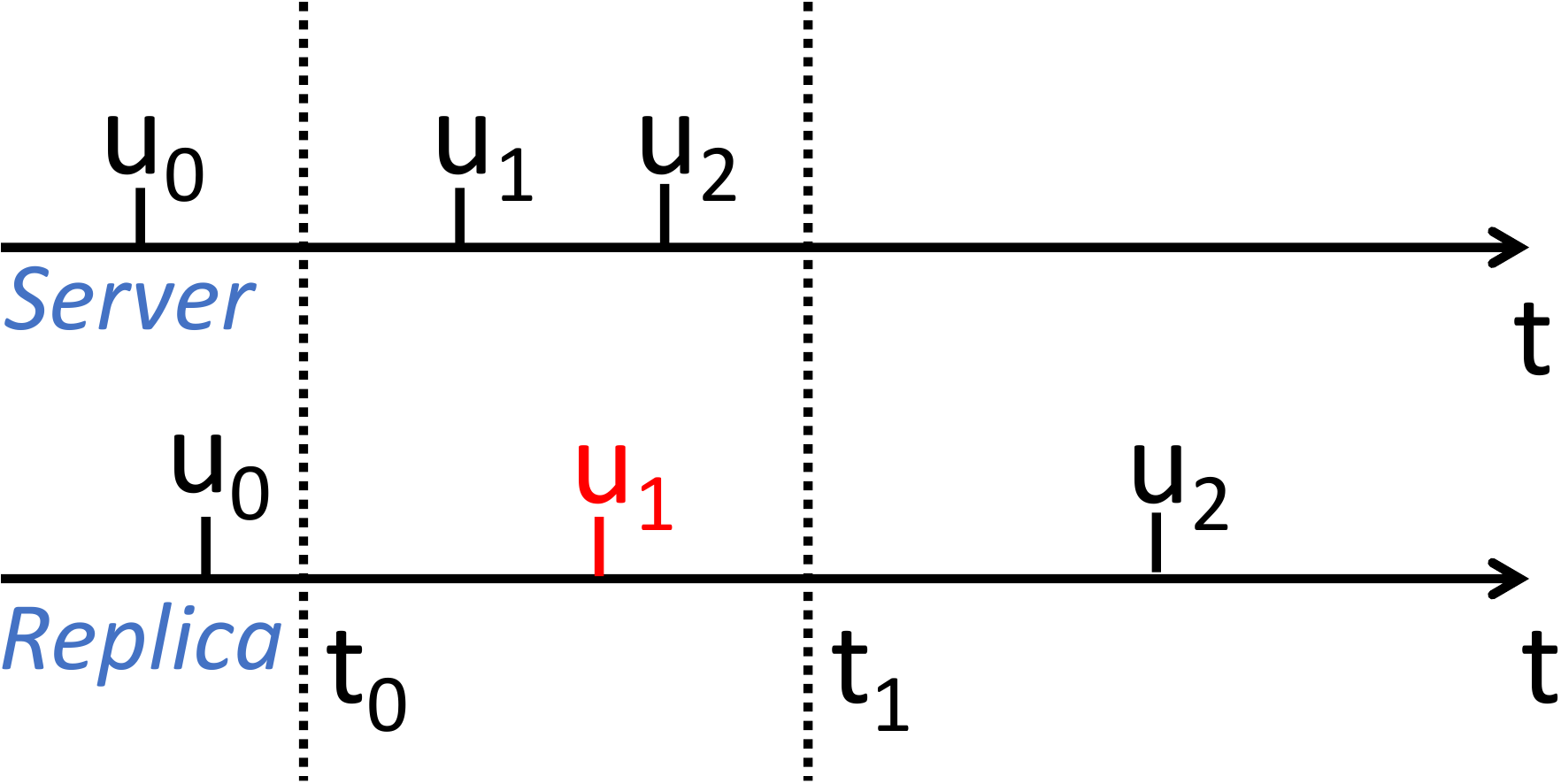}
      \subcaption{
        $u_1$ scheduled ahead at replica to satisfy divergence bound
        \label{fig:update-schedule-b}
      }
    \end{boxedminipage}%%
  \end{tabular}
  \caption{Update transfer schedule at server and replica}
  % \vspace{-0.1in}
\end{figure}%}}}

\noindent
\textbf{Enforcing bounded consistency:} Using in-network control, we
can ensure that all updates to the model at the server are also
applied to model at the replica in the exact same order. If the same
updates are always applied at the server and replica, then their models are
identical, realizing strict consistency. However, we can use
in-network control to {\em bound} model-replica divergence, thereby
leading to a flexible new notion of {\em bounded consistency}: here,
we simply ensure that $||\wb_s - \wb_r||_2$ is within a user-specified
bound, $Div_{max}$. Bounded consistency flexibly trades off the cost
of recovery against network efficiency of updates to primary and
replica models. Specifically, a large $Div_{max}$ bound allows
{\em delaying} several replicated updates, which can be aggressively
aggregated later, controlling network replication load.  A small bound
makes recovery fast but at the cost of higher replication load.

In-network control enables us to carefully schedule both
original and replicated updates to achieve bounded consistency. We now
show how in-network control can {\em reduce} divergence, which we
use in our replication algorithm
(\secref{sec:method-multiple-replicas}),

Consider a scenario where at time $t=0$, update $u_0$ is the latest to
be applied to the model at both the server and replica; the divergence
at $t=t_0$ is zero.  Now, consider a schedule of future updates
$u_1, u_2$ as shown in fig.~\ref{fig:update-schedule-a}. At time
$t=t_1$, the server leads the replica by two updates. By applying
eqn.~\ref{eq:async-sgd-server-compute} twice, the model divergence at
$t=t_1$ can be computed as:
\begin{small}
\begin{equation}
  || w^s_2 - w_0||_2 = ||(\gamma + \gamma^2) h^0 + (1+\gamma) u_1 + u_2||_2
  \label{eq:actual-divergence-two-updates}
\end{equation}
\end{small}
where, $h^0 = w_0 - w^{-1}$ is the update history at time $t=t_0$.

If $|| w^s_2 - w_0|| > Div_{max}$, then we can alter the schedule to
reduce the server's lead over the replica (e.g.,
fig.~\ref{fig:update-schedule-b}).  With the example new schedule,
divergence at time $t=t_1$
\begin{small}
\begin{equation}
  || w^s_2 - w^s_1||_2 = ||\gamma^2 h^0 + \gamma u_1 + u_2||_2
\end{equation}
\end{small}
is lower than divergence without network control. Our replication
algorithm in \secref{sec:method-multiple-replicas} uses this lead reduction idea.

% !TEX root = main.tex

\section{Architecture and APIs}
\label{sec:architecture}

\noindent
\textbf{Architecture:} The main component of \name is a
\emph{scheduler} that interacts with \name \emph{daemons} on each
worker/server; the scheduler processes update and model transfer requests from
the daemons and determines the (a) {\em next hop}, and (b) {\em schedule} for
each transfer.  The next hop can either be a final destination (worker or
server) or an intermediate \emph{aggregator} deployed alongside workers;
aggregators compute the {\em (weighted) sum} of incoming updates and forward
them to the next hop determined by the scheduler.  A \emph{network monitor}
periodically measures and reports available network bandwidth to the scheduler
which is used to make scheduling decisions.  \name daemons are responsible for
interfacing with application entities using \sysname{} APIs and enforcing the
scheduler's decisions.

\lstset{
  language=Python,
  basicstyle={\tiny\ttfamily},
  numbers=left,
  escapeinside={<@}{@>},
}

\begin{table}[t]%{{{
  \scriptsize
  \centering
  \begin{tabular}{|c|l|} \hline
    & \textbf{MLFabric APIs} \\ \hline
    & \texttt{registerAsWorker(\textit{params})} \\
    worker & \texttt{push(\textit{server}, \textit{update}, {\color{red}\textit{update\_norm}})}\\
    & \texttt{get(\textit{server}, \textit{model})} \\
    % & \texttt{AllReduce(\textit{update})} \\
%%     & \texttt{clock(\textit{iteration})} \\
    & {\color{red}\texttt{AllReduce(update)}} \\
    \hline
    & \texttt{registerAsServer(\textit{params})} \\
    server & \texttt{registerUpdateCallback()} \\
    & \texttt{registerRequestCallback()} \\ \hline
    replica & {\color{red}\texttt{registerAsReplica(\textit{server}, \textit{params})}}\\
    & {\color{red}\texttt{registerUpdateCallback()}} \\ \hline
%%     \texttt{\textit{params}} & \textit{consistency := synchronous | asynchrorous} \\
    \texttt{\textit{params}} & {\color{red}\textit{delay bound :=  $\tau_{max}$}}\\
    & {\color{red}\textit{divergence bound :=  $Div_{max}$}}\\ \hline
  \end{tabular}
  \caption{Items in red are extensions we make to the PS API.\label{tab:mlfabric-apis}}
%%   \vspace{-0.2in}
  % \vspace{-0.1in}
\end{table}%}}}

\noindent
\textbf{APIs:} \sysname{} extends existing PS APIs (see
\tabref{tab:mlfabric-apis}). It also provides an MPI AllReduce API which is
realized through PS APIs (\secref{sec:other-aspects}).
% ; we extend the communication APIs in Bosen~\cite{bosen}.
These APIs help realize the optimizations discussed in \xref{sec:motivation}.
For example, for bounded consistency replication
(\secref{sec:in-network-replication}), \sysname{} allows machines to register
as replicas and allows workers to specify the norm of the update when it is
pushed.

% !TEX root = main.tex
\section{Algorithms}
\label{sec:method}

\sysname{} scheduler determines the communication pattern for a batch
of updates available from workers. It computes the {\em transfer
  schedule} (i.e., how bytes in an update are transferred at any given
time) and {\em forwarding} (next hop -- i.e., server or intermediate
aggregator hop) for each of these updates.  This is done so as to
(1) minimize the average completion time of update transfers
(\xref{sec:time-sharing}), improve network efficiency
(\xref{sec:aggr}), and make fresh models available earlier
(\xref{sec:time-sharing} and \xref{sec:aggr}), while (2) bounding
worst-case delay (\xref{sec:theorem}) even under stragglers or
changing network conditions. Also, when replica servers are
deployed, \sysname{} schedules minimal replication traffic to bound
primary-replica model divergence (\xref{sec:in-network-replication}).

We first formulate an integer linear program (ILP) to jointly
determine the optimal schedules of updates and forwarding for
aggregation (\xref{sec:one-shot-optimization}). But the ILP is
intractable even for a PS-system with one server and no replicas. It is intractable even when determining the schedules alone
(i.e., aggregation is also ignored) while considering delay
bounds.

To handle this intractability, we decompose the problem and solve it
progressively. As mentioned earlier, we process a batch of updates at
a time. We first determine {\em ordering} for the batch of
updates. Second, given ordering, we determine the {\em
  forwarding/aggregation strategy}, which results in {\em tentative
  schedules} for transferring updates to either servers or
aggregators. Third, given the ordering, and schedules of updates, we
determine which {\em replica transfers} to schedule and when so that they
finish before updates in the batch are committed at the server; if
this replication ``falls short'', i.e., causes server-replica
model divergence, we delay a small number of the tentative primary
server transfer schedules such that divergence bound is met. In the end, we
have {\em concrete transfer schedules} for all updates in the batch
and for those that are replicated. For simplicity we assume a single
server $S$.\footnote{In \secref{sec:method-multiple-servers}, we
  consider the case where the model is sharded across multiple
  parameter servers.} We describe the above three algorithms in turn.

\newcommand{\sjfalgorithm}{%
\begingroup
\removelatexerror
\begin{algorithm}[H]
  \scriptsize
  \DontPrintSemicolon
  \SetInd{0em}{0.5em}
  \SetKwInOut{Input}{In}\SetKwInOut{Output}{Out}
  \Input{$U$ (Updates), $NW$ (Network state), $S$ (server)}
  \Output{$\mathcal{O}(U)$ (Ordered updates)}
  \SetKwFunction{ShortestUpdate}{ShrtUp}
  \SetKwProg{FDeadline}{Fn}{}{\KwRet $g^*$}
  \FDeadline{\ShortestUpdate{$UU$, $NW$}}{
    \tcp{$t_{en}$: As described in fig.~\ref{fig:end-time-calculation}}
    $g^* \gets \argmin\limits_{g \in UU}~t_{en}(g, NW, S)$\;
  }
  \SetKwFunction{NetworkUpdate}{NetUp}
  $\mathcal{O}(U) \gets [\quad]$\;
  \For{$i \gets 1$ \KwTo $|U|$}{
    $g^* \gets~$\ShortestUpdate{$U - \mathcal{O}(U)$, $NW$}\;
    $\mathcal{O}(U) \gets [\mathcal{O}(U), \quad g^*]$\;
    \tcp{\NetworkUpdate: See fig.~\ref{fig:network-update}}
    $NW \gets~$\NetworkUpdate$(NW, g^*, S)$ \label{line:network-update}
  }
  \caption{\scriptsize\label{code:sjf-ordering}}
  %\vspace{-0.1in}
\end{algorithm}
\endgroup
}%

\begin{figure}
  \centering
  \begin{tabular}{@{}c@{}c@{}}
    \multirowcell{2}{%
      \begin{boxedminipage}[!t]{0.5\linewidth}
        \vspace{-0.7in}
        \centering
        \sjfalgorithm
        \subcaption{\label{fig:sjf-ordering-wrapper}}
      \end{boxedminipage}%
    }%
    &%
    \begin{boxedminipage}[h]{0.5\linewidth}
      \includegraphics[width=\linewidth]{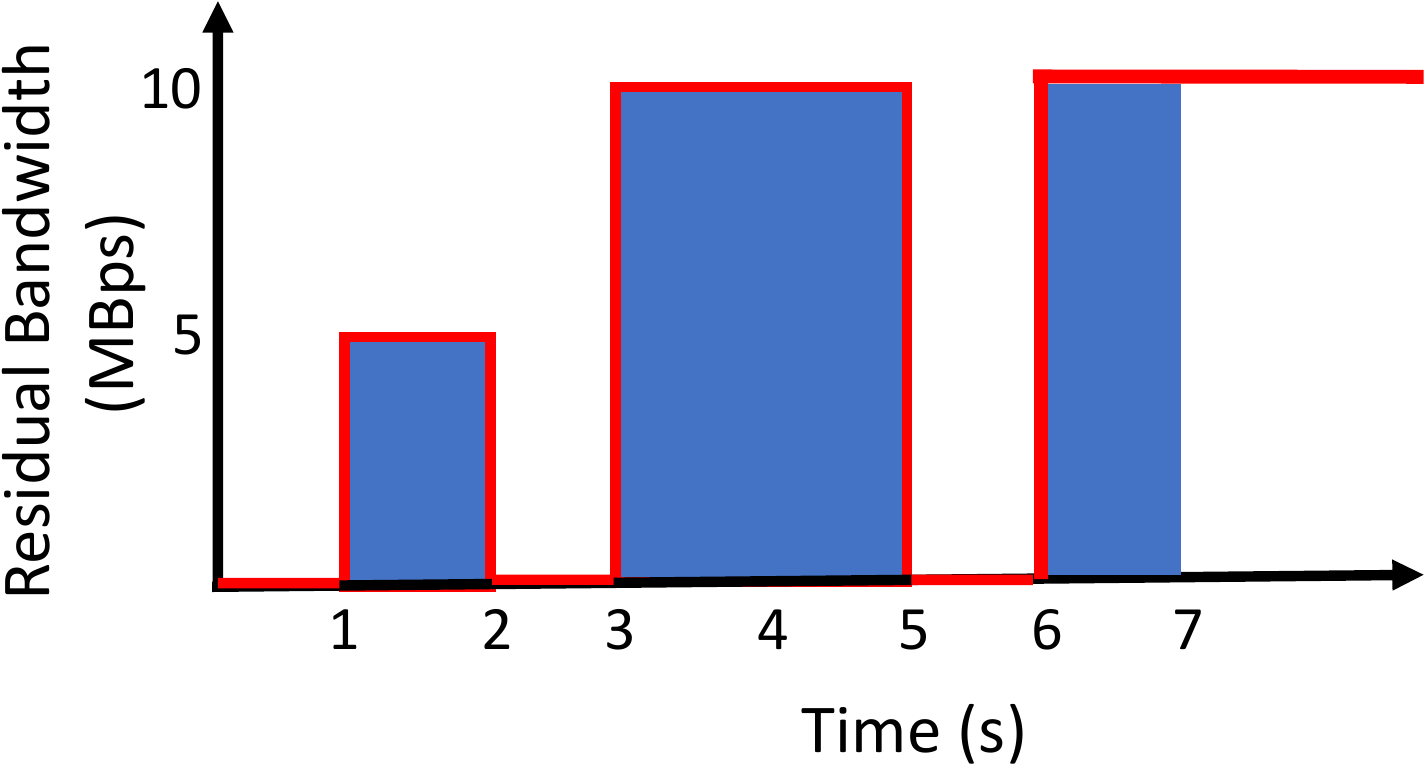}
      \subcaption{\label{fig:end-time-calculation}}
    \end{boxedminipage} \\%%
    &%%
    % \hspace{0.02\linewidth}%%
    \begin{boxedminipage}[b]{0.5\linewidth}
      \includegraphics[width=\linewidth]{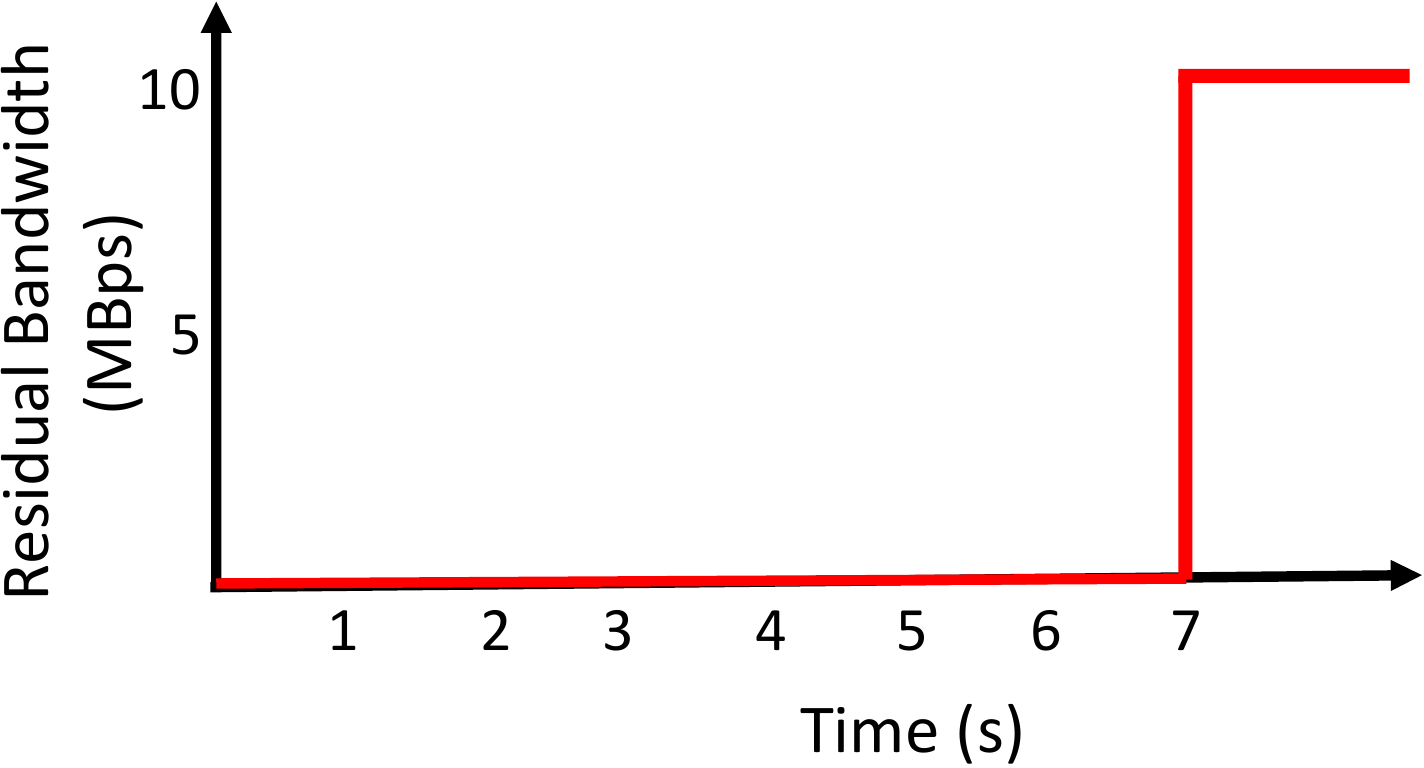}
      \subcaption{\label{fig:network-update}}
    \end{boxedminipage}
  \end{tabular}
  \caption{Ordering available updates. (a) Shortest transfer first ordering pseudo-code.
  (b)$t_{en}$ calculation. Consider an update, $g$, of size $30$~MB, available
  at time $t=0$. The red line represents residual bandwidth along the path for
  $g$. The blue shaded region represents the bandwidth utilized by update, $g$.
  Here, $t_{en}(g)=7$. (c) Network b/w update. Residual bandwidth after reserving bandwidth for $g$.}
  % \vspace{-0.1in}
\end{figure}

\subsection{Update Ordering}
\label{sec:method-ordering-updates}

Given a set of available worker updates ($U$), and a single
server, we first describe how we determine the order
($\mathcal{O}(U)$) in which updates are transferred over the
network. We ignore replication/aggregation for now.

We assume network time-sharing (\secref{sec:time-sharing}), i.e.,
updates transferred on a bottleneck link do not have overlapping
transfer times. Given this, we attempt to determine an ordering that
(1) minimizes the average update transfer time to the server to ensure
a fast rate of updates (\secref{sec:re-ordering}), (2) subject to the
constraints that delays bounds are met. Since this problem in itself
is also intractable, we develop a heuristic that decouples them by
first attempting to minimize average transfer time (\secref{sec:avg}),
and then ``fixing'' any violated delay bounds
(\secref{sec:method-bounding-delay}). This heuristic may result in
network links/server NIC laying ``fallow'', and we show how to alter
the ordering to address this inefficiency without violating delay
bounds (\secref{sec:method-dropping-updates}).

\begin{figure}
\centering
\includegraphics[width=0.44\linewidth]{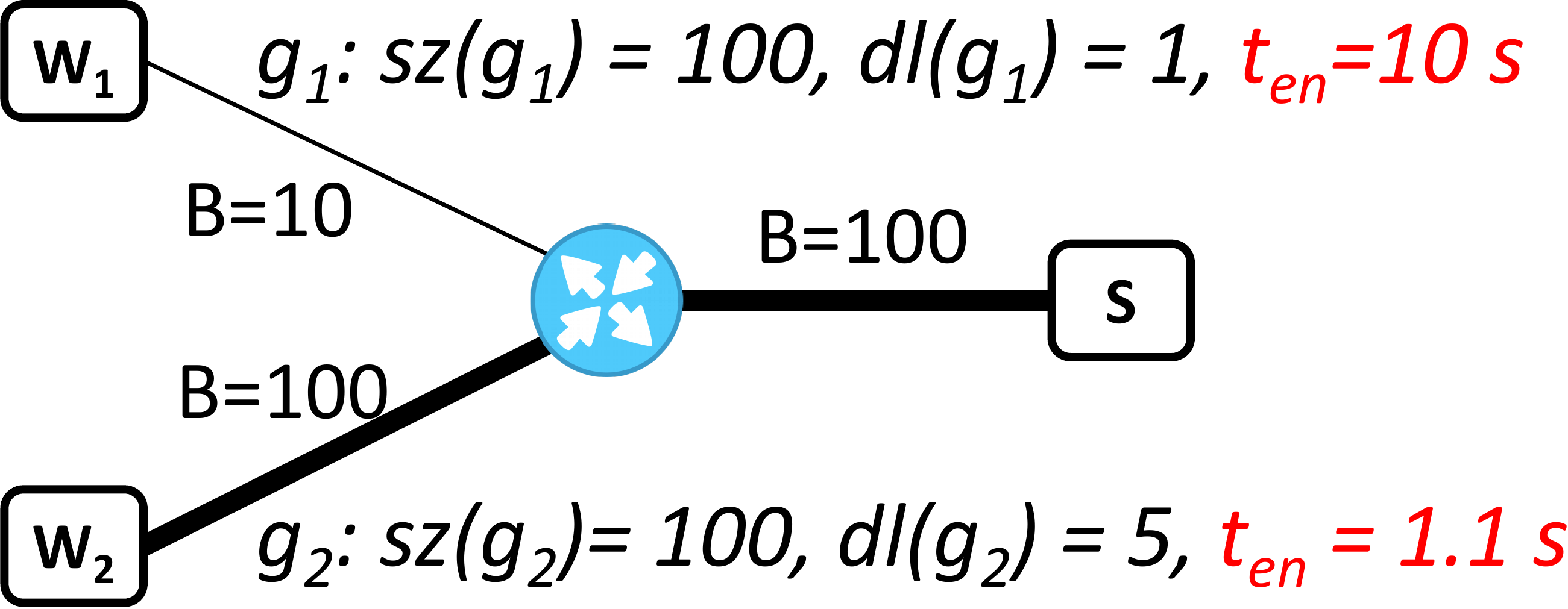}
\caption{A case for preemptively dropping updates. Update $g_1$ takes $10$ s to
  complete because of low bandwidth behind worker $w_1$.}
\label{fig:preemptive}
  % \vspace{-0.1in}
\end{figure}

\subsubsection{Average completion time}
\label{sec:avg}

To determine an order that minimizes average update transfer time, we
iteratively emulate {\em shortest-job-first} ordering for update transfers
(alg.~\ref{code:sjf-ordering}): in each iteration, given current
available bandwidth, we compute each single update's transfer
completion time, $t_{en}$, by factoring in the bottleneck bandwidth
the transfer has available over time, and determining how the bytes in
the update are transferred by maximally using bottleneck capacity at
any time (fig.~\ref{fig:end-time-calculation}). We pick the transfer
with least completion time, and reserve capacity on its path over time; the amount of reservation equals the time-varying bottleneck
bandwidth, and reservation duration equals the transfer completion time. We
then update remaining network capacities over time, and iterate
(line~\ref{line:network-update}, alg.\ref{code:sjf-ordering}).

\subsubsection{Bounding delays}
\label{sec:method-bounding-delay}

Shortest-transfer-first-ordering can increase delay (potentially
greater than the configured upper bound, $\tau_{max}$) for large
updates or those with less bandwidth to the server.  To ensure that
these are transferred earlier in the order so as to meet delay bounds,
we introduce transfer \emph{deadlines}.  For an ordering
$\mathcal{O}(U)$ over updates in $U$, the deadline for update $g \in
U$ is:
\begin{small}
\begin{equation}
  dl(g) := v(g) + \tau_{max} - v_{init}
\end{equation}
\end{small}
where, $v(g)$ is the model version for update $g$ and $v_{init}$ is the version
of the model after all updates from previous batchs are applied.

We then modify the update ordering algorithm
(alg.~\ref{code:sjf-ordering}) as follows: in iteration $i$ if there
exists an unscheduled $g \in U$, such that $dl(g) = i$, then we pick
$g$ in that iteration and reserve bandwidth for
transferring $g$ as above; otherwise we greedily pick the update with the least
transfer time ($t_{en}$).

\subsubsection{Dropping delayed updates}
\label{sec:method-dropping-updates}

Unfortunately, simply accounting for shortest-transfer-first and
deadlines does not suffice in determining a ``good'' ordering. Unless
care is taken while factoring deadlines, the ordering may
unnecessarily lead to network or server resources staying fallow. To
see why, consider two workers $w_1$ and $w_2$ with updates $g_1$ and
$g_2$. Let the deadline for the two updates be $dl(g_1) = 1$ and
$dl(g_2) = 5$. Let the network topology and current state of the
network be as shown in \cref{fig:preemptive}.
Since $g_1$ has a deadline of 1, the above  approach
picks $g_1$ as the first update to apply to the model, and thus
transfers it first. Because its bottleneck bandwidth is 10Mbps, the
transfer would take 10s. In the next iteration, the algorithm
selects $g_2$. If $g_2$ is scheduled immediately, then its available
bandwidth is 90Mbps (after bandwidth for $g_1$ is reserved), and the
update takes 1.1s to finish. Thus, $t_{en}(g_1) > t_{en}(g_2)$, which
violates $g_1$'s delay bound (recall $dl(g_1) = 1$). One way to avoid
this is to transfer $g_2$ in a delayed manner such that $g_1$ is
applied first, but this leaves 90Mbps of network capacity on the link
to the server unused while only $g_1$ is being
transferred. Alternately, $g_2$ can be transferred per the above
ordering, but applied only after $g_1$ is applied -- in this case the
server stays idle while waiting for $g_1$ even though $g_2$ is
available to be applied.

To ensure work conserving server and network behavior as well as delay
bounds, we modify our algorithm to {\em drop} the update $g_1$ at the
worker itself. $g_2$ is then immediately scheduled for transfer.

\newcommand{\orderingalgorithm}{%
\begingroup
\removelatexerror
\begin{algorithm}[t]
  \scriptsize
  \DontPrintSemicolon
  \SetInd{0em}{0.5em}
  \SetKwInOut{Input}{In}\SetKwInOut{Output}{Out}
  \Input{$U$ (Available updates), $NW$ (Network state), $S$ (server)}
  \Output{$\mathcal{O}(U)$ (Ordered updates)}
  \SetKwFunction{ShortestDeadline}{ShrtDline}
  \SetKwProg{FShortest}{Fn}{}{\KwRet $g^*$}
  \FShortest{\ShortestDeadline{$it$, $UU$, $NW$}}{
    $g^* \gets~$ \lIf{$\exists g : dl(g) = it$}{
    $g$
    } \lElse {
    \ShortestUpdate{$UU$,$NW$}
    }
  }
  \SetKwFunction{NetworkUpdate}{NetUp}
  % $\mathcal{O}(U) \gets [\quad]$\ ;
  $P \gets \phi$ \tcp*{processed updates}
  \lFor{$i \gets 1$ \KwTo $|U|$}{
  $dl(g_i) = v(g_i) + \tau_{max} - v_{init}$
  }
  \For{$i \gets 1$ \KwTo $|U|$}{
    $g^* \gets~$ \ShortestDeadline$(i$, $U - P$, $NW)$\;
    $P \gets~P~\cup~g^*$\;
    $g^{\circ} \gets~$ \ShortestDeadline$(i+1$, $U - P$, \NetworkUpdate$(NW, g^*, S))$\;
    \lIf{$t_{en}(g^*, NW, S) > t_{en}(g^{\circ},$ \NetworkUpdate $(NW, g^*, S), S)$}{
    \textbf{continue}
    }
    $\mathcal{O}(U) \gets [\mathcal{O}(U), \quad g^*]$\;
    $NW \gets~$\NetworkUpdate$(NW, g^*, S)$
  }
  \caption{\footnotesize Final update ordering algorithm \label{code:final-ordering-algorithm}}
\end{algorithm}
\endgroup
}%

\begin{centering}
  \orderingalgorithm
\end{centering}

Thus, our iterative algorithm requires the following fix: in every
iteration of the modified algorithm from
\secref{sec:method-bounding-delay}, where we pick an update (call it
``current'') to satisfy a deadline, we {\em look-ahead} and determine
the completion time of the next update that will be applied (call it
``next'').  If ``next'' completes before ``current'', we discard
``current''. The final ordering algorithm that combines
shortest-job-first, meets delay bounds, and avoids wasting resources is in Alg.~\ref{code:final-ordering-algorithm}.

\begin{figure}[!t]
  \centering
  \includegraphics[width=0.75\linewidth]{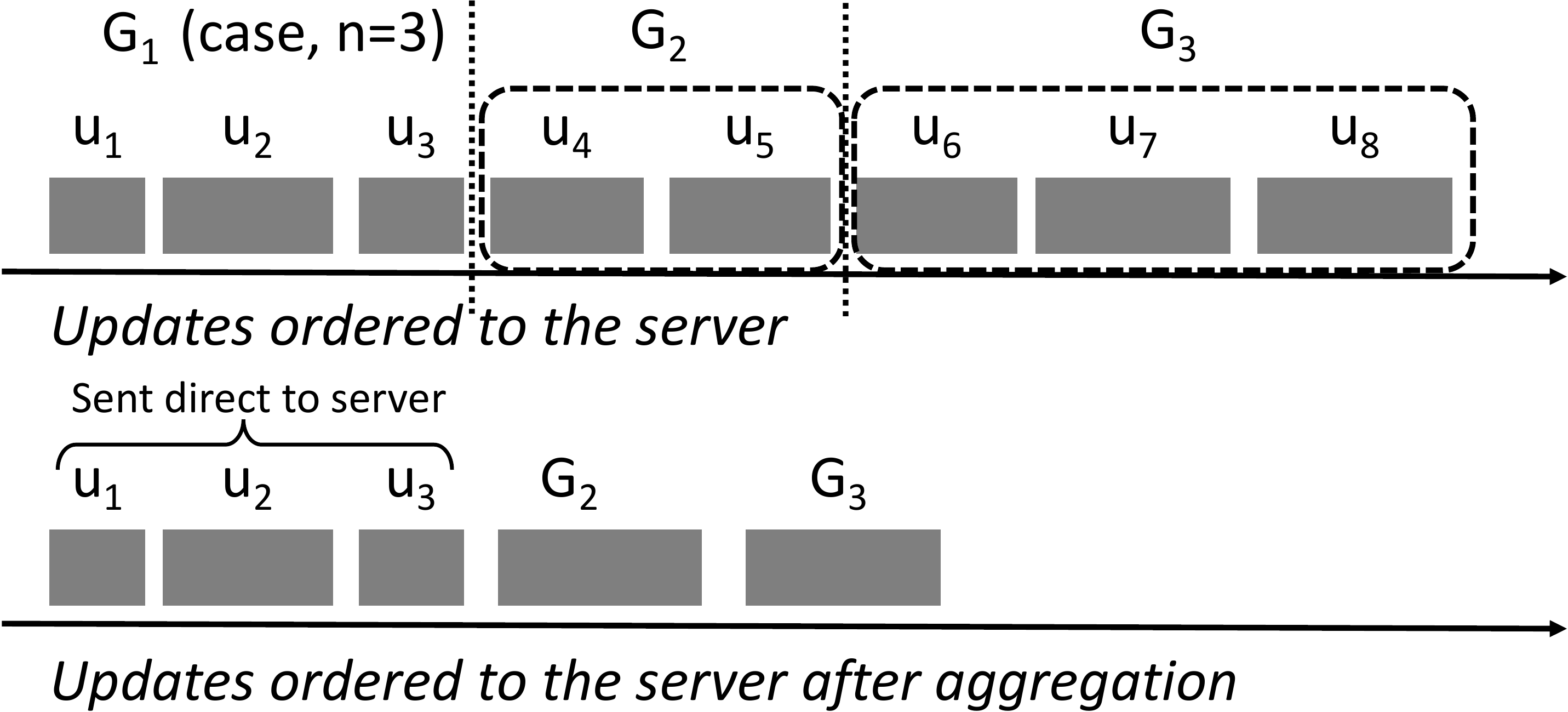}
  \caption{Partitioning ordered updates to server. Later partitions are
    aggregated before being sent to server. $G_i$ are the groups. The figure
    depicts the case where first 3 updates are sent directly to the server. Note that
    $u_6$ is not added to $G_2$ since time taken to aggregate $u_4, u_5, u_6$
    would exceed the time taken to send $u_1, u_2, u_3$ to the server.
  }
  \label{fig:chunking}
  \vspace{-0.1in}
\end{figure}

\subsection{Aggregation}
\label{sec:in-network-aggregation}

With the transfer order determined, we describe how to
opportunistically aggregate updates in-network. The goal is to use
spare compute and network capacity at non-server (aggregator)
locations to aggregate as efficiently as possible while
preserving the above-determined ordering.

We achieve this by grouping ordered updates in a clever manner, and
streaming each group to either the server directly, or first to an
aggregator and then the server, such that the server always has a
constant stream of ordered or ordered+aggregated updates arriving at
its NIC (\cref{fig:chunking}). In some more detail, given a set of $k$
aggregators for the server, we partition ordered updates to the server
into $k+1$ groups, using an algorithm we describe shortly. Given the
partitioning, the first of these groups, if non-zero in size, is
forwarded directly to the server.  All gradients in subsequent groups
are aggregated before they are forwarded to the server
(\cref{fig:chunking}). Further, updates in each group are forwarded to
aggregators as per $\mathcal{O}(U)$ determined above. Thus we ensure
that, (a) the delay constraints remain satisfied
(\cref{fig:chunking}), and (b) update to the model is consistent to
the case with no aggregation. The output from the aggregator also
obeys the same order.

Our algorithm for determining the best way to partition updates into
$k+1$ groups for the server is key to {\em efficiency} and is shown in
Alg.~\ref{code:aggregation-algorithm}.  The algorithm determines group
membership so as to minimize the total time until the aggregated
update from the last $(k+1)^{st}$ group is transferred to the server.
The partitioning is guided by the following key {\em constraint}:
aggregating \emph{all} updates in the $i^{th}$ group from the
corresponding workers should not finish later than the time when all
prior $i-1$ groups' gradient aggregates are transferred to the server.
This condition ensures efficiency, i.e., the server NIC is never left
fallow, waiting for updates to be aggregated.

We first randomly pre-assign the aggregator to use for the $i^{th}$
group. Then, the aggregation algorithm works are follows. Given $|U|$
ordered updates, we exhaustively enumerate $|U|+1$ cases (lines
21-23). In the $n^{th}$ such case, $n= 0,\ldots,|U|$: (1) the first $n$
updates are forwarded directly to the server (lines 3-7). (2) We
greedily assign successive updates to the first aggregator as
long as the above-mentioned constraint is satisfied (lines 16-18). (3)
When the constraint is violated, we greedily start assigning updates
to the second aggregator (lines 10-15), and so on. \figref{fig:chunking} shows this for $n=3$.

\begin{centering}
  \begin{algorithm}[!t]
    \scriptsize
    \DontPrintSemicolon
    \SetInd{0em}{0.5em}
    \SetKwInOut{Input}{Input}\SetKwInOut{Output}{Output}
    \Input{$\mathcal{O}(U)$ (ordered updates), $NW$ (network), $S$ (server)}
    \Output{$A(g), \forall g \in \mathcal{O}(U)$ (aggregator for update, $g$)}
    \SetKwFunction{Aggregate}{DetAgg}
    \SetKwProg{FAgg}{Fn}{}{\KwRet $t_{max}$}
%%     \For{$i~\gets~1$ \KwTo $|U|$}{
%%      \Aggregate$(i, \mathcal{O}(U), NW, A)$ }
    \FAgg{\Aggregate$(n, \mathcal{O}(U), NW, A)$}{
      $aid~\gets~0,\quad t_{max} = 0$ \tcp*{$aid = 0$ indicates fwding to server}
      \For{$i~\gets~1$ \KwTo $n$}{
        $A(g_i)~\gets~aid$\;
        $t_{max}~\gets~t_{en}(g_i, NW, S)$\tcp*{Update time until server is blocked}
        $NW~\gets~$\NetworkUpdate$(NW, g_i, S)$\tcp*{Reserve bandwidth for direct transfer to server}
      }
      $aid~\gets~aid+1$,~ $i~\gets~n+1$\;
      \While{$i \leq |\mathcal{O}(U)|$}{
        \If{$t_{en}(g_i, NW, aid) > t_{max}$}{
          $t_{max}~\gets~t_{en}(a_{aid}, NW, S)$\tcp*{$a_{aid}$: aggregated update at $aid$}
          $NW~\gets~$\NetworkUpdate$(a_{aid}, NW, S)$ \tcp*{Reserve bandwidth from $aid$ to server}
          $aid~\gets~aid+1$\;
          \textbf{continue}
        }
        $A(g_i)~\gets~aid$\;
        $NW~\gets~$\NetworkUpdate$(g_i, NW, aid)$\tcp*{Reserve bandwidth to aggregator, $aid$}
        $i~\gets~i+1$\tcp*{Consider next update in order}
      }
    }
    \tcc{Enumerate all cases and store aggregation pattern and total time.}
    \For{$i~\gets~0$ \KwTo $|\mathcal{O}(U)|$}{
      \Aggregate$(i, \mathcal{O}(U), NW, A)$
    }
    $i^*~\gets~\argmin\limits_{i \in [0, |\mathcal{O}(U)|]}$ \Aggregate$(i, \mathcal{O}(U), NW, A)$
    \caption{\footnotesize Aggregation algorithm.\label{code:aggregation-algorithm}}
  \end{algorithm}
  % \vspace{-0.25in}
\end{centering}

After every assignment of an update to server/aggregator, we
reserve network bandwidth for the transfer (lines 5, 6, 17). We also
reserve bandwidth for transferring the aggregated update from the
aggregator to the server (lines 11, 12).

All $|U|+1$ cases result in a different aggregation patterns. We pick
the one which takes the least amount of time to transfer all the
updates to the server (line 24).

Note that our algorithm does not alter the transfer schedules
for updates in group 1 compared to those computed by
Alg.~\ref{code:final-ordering-algorithm}.  For all other updates, and
aggregates, our algorithm computes new transfer schedules that differ
from Alg.~\ref{code:final-ordering-algorithm}, because the schedule
now accounts for new transfer destinations (aggregators vs. the
server) and new transfers/sources (aggregates from aggregators
vs. original updates from workers).

We now have transfer schedules for a batch $U$ where: delay bounds are
met; updates are aggregated efficiently; updates/aggregates are
committed as fast as possible to the server ensuring high model update
rate; and server NIC and overall network efficiency are high.

\subsection{Replication}
\label{sec:method-multiple-replicas}

Given the above transfer schedules for  a
batch, we now describe our replication algorithm. It determines the final schedules for transfers to both the
server and the replica; thus, we refer to the above-determined
schedules (alg.~\ref{code:final-ordering-algorithm}) as ``tentative''.

Suppose there is just one replica (extending to more replicas is
simple).  The goal is to transfer a prefix of the $|U|$
updates to replica in the {\em same order} as $\mathcal{O}(U)$
(determined above), such that
when all the updates in the batch are committed at the server the
divergence bound is satisfied.
%% replicating a transfer subset ensures
%% network efficiency of replication.
We assume a separate $k'$
aggregators are earmarked for the replica.

The replication algorithm operates in cognizance of the above
``tentative server schedules'' and the resulting network state.
It first computes ``tentative replica schedules''\footnote{Note:
  replicated transfers use already-computed server transfer ordering}
using the aggregation alg.~\ref{code:aggregation-algorithm}, where the
initial state of the network accounts for tentative server (original
transfer) schedules.

Suppose the last transfer to finish in the tentative server schedules
finished (i.e., committed at the server) at de
$T_{last}$. We check if the divergence bound, $Div_{max}$, holds
at $T_{last}$ based on the server and replica updates that would
have been committed by this time; we can determine this from tentative
server and replica schedules as shown shortly. Note that only a
prefix of $U$ server updates would have completed by $T_{last}$.

If the bound is satisfied, we freeze the replica schedule (i.e., apply
to replica) for all worker-to-aggregator and aggregator-to-replica
transfers that would have finished by $T_{last}$; all replicated
updates that finish after $T_{last}$ are ``punted'' to be processed
along with the next batch.

If the divergence bound is not satisfied at $T_{last}$, then we delay (akin to the example in (\xref{sec:in-network-replication})
{\em just the last update}  in the tentative server schedule to start
after completion of the earliest update in the replica
schedule (say, $a_{e}$) such that the divergence bound is satisfied;
all replica updates until $a_{e}$ are then frozen (this is still a
prefix of $U$); subsequent updates are punted to the next batch.

Punting the processing of some replicated updates
to the next batch has two advantages for said next batch: (1) punted
transfers combine with the set of replicated updates for
said batch, which helps increases overall aggregation efficiency in
transfers to the replica when said batch is
processed\footnote{especially when bottleneck is close to replica
  server}, and, (2) in turn, this helps free up resources for server
update transfers in said batch, helping them to finish faster. With a
large divergence bound, more such updates are punted to subsequent
batches, magnifying the aforementioned benefits. We show this
empirically in \secref{sec:evaluation}.

Thus, we now have {\em concrete} schedules for both server and
replica updates.

We return to the problem of efficiently estimating the divergence
between sets of updates in the tentative server and replica
schedules. Because computing exact divergence requires expensive computation
over large updates, in \sysname{}, we approximate
actual divergence by an upper bound.
Suppose at any time, the latest updates to the server and replica are, $w^2_s$
and $w^0$ respectively (see \cref{eq:actual-divergence-two-updates}).
The divergence  can be approximated using Cauchy-Schwartz inequality as:
\begin{small}
\begin{eqnarray}
  || w^2_s - w^0||_2 &=& ||(\gamma + \gamma^2) h^0 + (1+\gamma) u_1 + u_2||_2 \\
  &\leq& \{ a_1 ||h^0||^2_2 + a_2 ||u_1||^2_2 + a_3 ||u_2||^2_2  \nonumber \\
  && a_4 ||h^0||.||u_1|| + a_5 ||u_1||.||u_2|| \nonumber \\
  &&  + a_6 ||u_1||.||h^0|| \}^{\frac{1}{2}} \nonumber \\
  &=& ||w_s^2-w^0||_2^{approx.} \label{eq:divergence-ub}
\end{eqnarray}
\end{small}
where, $h^0$ is the update history and $a_1, \ldots, a_6$ are
constants dependent on momentum $\gamma$. The value
$||w_s^2-w^0||_2^{approx.}$ (denoted as $div(s,r)$) can be easily
computed with just the norm of the individual updates provided by the workers/server to the \sysname{}
scheduler (\secref{sec:architecture}); verifying that the
approximation is less than $Div_{max}$ is sufficient for bounded
consistency.

% !TEX root = main.tex
\section{Extending \sysname{}}
\label{sec:other-aspects}

We now describe how \sysname{} applies to synchronous and stale synchronous SGD, and to MPI frameworks.

\noindent
\textbf{Synchronous SGD/PS:}
Here, at each iteration, workers read the latest model and compute
a local update using a portion of the mini-batch. The updates are then
aggregated at the server and applied to the model (also incrementing model
version) before the start of next iteration.
\sysname{}'s approach to construct {\em dynamic network-aware}
aggregation topologies naturally helps synchronous SGD.
The workers' updates for an iteration are batched when they become
ready, say into $|U|$ updates in a batch. Since update ordering does
not apply to synchronous SGD, aggregation here starts with a {\em
  list} of updates (vs. an ordering in
\secref{sec:in-network-aggregation}). Then, directly applying our
algorithm from \secref{sec:in-network-aggregation} ensures that this
batch of updates is transferred as efficiently as possible.  The next
batch may use a different aggregation topology.  Our replication
algorithm (\secref{sec:method-multiple-replicas}) also applies
directly. Note that we have to guarantee bounded divergence only
at the end of an iteration (after all workers' updates are applied) as
opposed to end of a batch.

\noindent {\bf Stale Synchronous SGD/PS:} Stale synchronous (SS) SGD
is a consistency model that allows slow workers to lag behind fast
workers by up to $K$ iterations~\cite{parameterserver-osdi}; typically
$K\sim 2$. This form of delay management in SS is restrictive
compared to the delay management that \sysname{} enables for
asynchronous SGD. For example, with $K=2$, the maximum staleness of a
model is bound by $2\times num\_workers$.  However, a worker that is
more than twice slower than other workers will halt progress of all
other workers until the slow worker progresses
to the next iteration.  In contrast, a delay bound of
$2\times num\_workers$ in asynchronous SGD with \sysname{} will not halt
other workers' progress, while at the same time ensuring the staleness
of the update is less than $2\times num\_workers$.

Further, a typical implementation of SS does not implement update
aggregation. In-network control offered by \sysname{} can be applied to
update aggregation for SS in a manner similar to synchronous/PS above.

\noindent \textbf{MPI:} \sysname{}'s \texttt{AllReduce}
API can be used by existing MPI-based systems to implement
synchronized SGD. Internally, \sysname{} would implement 
\texttt{AllReduce} through successive calls to: (a)
\texttt{push(\textit{root, update, norm})} and (b)
\texttt{get(\textit{root}, \textit{update})}, using a synchronous
consistency model; \texttt{\textit{root}} is randomly chosen among the
workers and acts as the root of the aggregation topology, which is a
dynamically constructed tree (similar to synchronous/PS
above). \texttt{get(\textit{root, update})} pulls the aggregated
update from \texttt{\textit{root}} once updates from all workers are
received.

\sysname{} can also help with model distribution (\secref{sec:method-model-distribution}).

% !TEX root = main.tex
\section{Evaluation}
\label{sec:evaluation}

\newcommand{\baseasync}{{\emph{Async}}\xspace}
\newcommand{\servasync}{{\emph{Async-S}}\xspace}
\newcommand{\delayctrlasync}{{\emph{Async-D}}\xspace}

\newcommand{\appzero}{{\emph{App-350K}}\xspace}
\newcommand{\apphundred}{{\emph{App-100M}}\xspace}
\newcommand{\apptwohundred}{{\emph{App-200M}}\xspace}

\newcommand{\simplesgd}{{\emph{Sync}}\xspace}
\newcommand{\treereduce}{{\emph{Tr-Sync}}\xspace}

\begin{figure*}[t]
  \centering
  \begin{tabular}{@{}c@{}c@{}c@{}c@{}}
    \multicolumn{4}{@{}l}{%%
      \centering
      \begin{boxedminipage}[b]{\linewidth}
        \centering
        \includegraphics[width=0.6\linewidth]{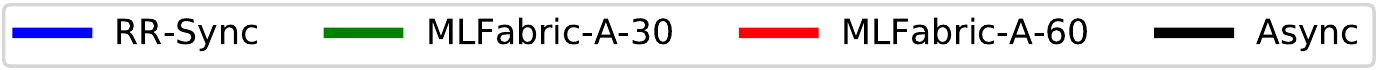}
      \end{boxedminipage}%%
    }
    \\%%
    \begin{boxedminipage}[b]{0.25\linewidth}%%
      \centering
      \includegraphics[width=0.9\linewidth]{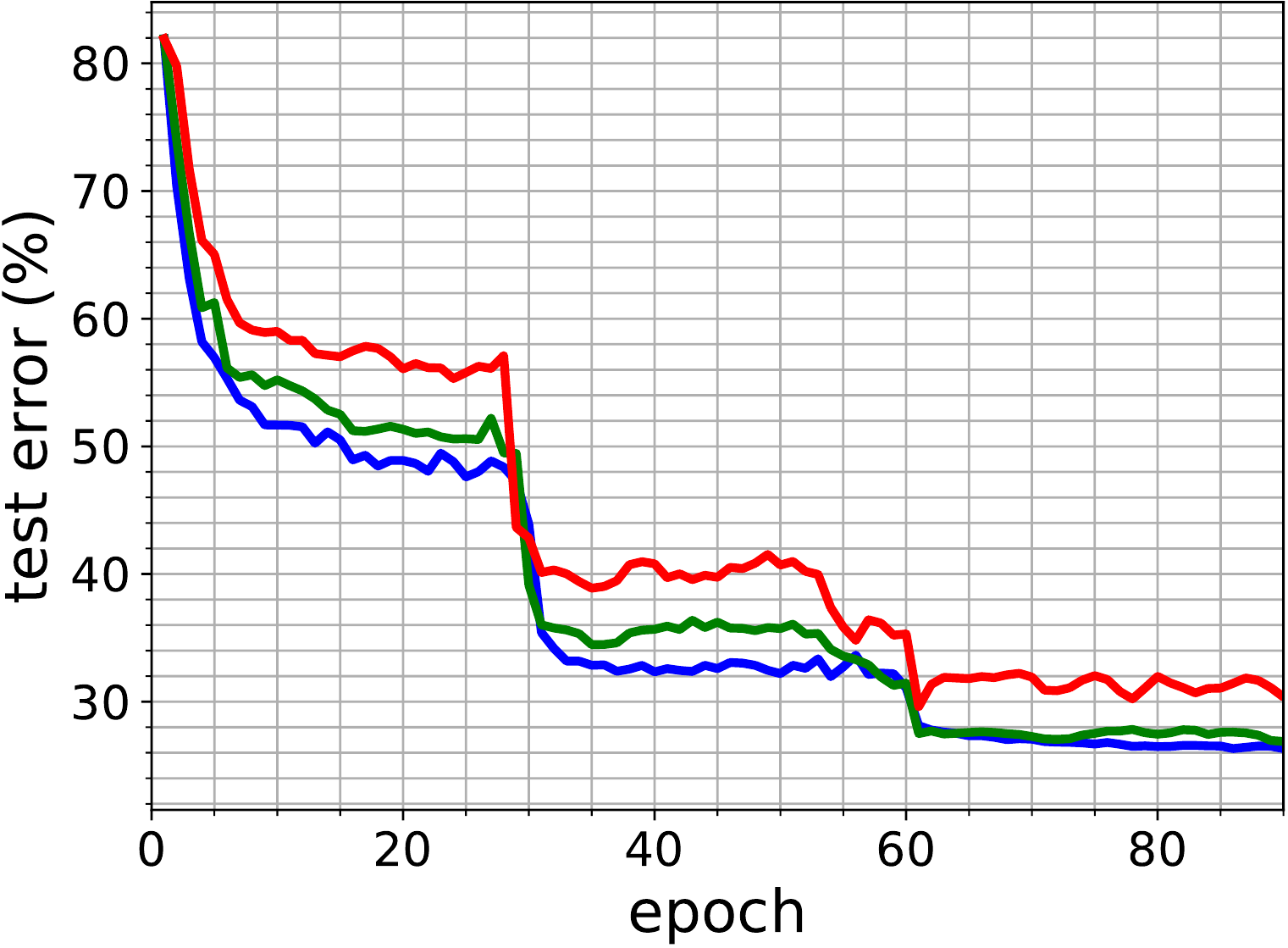}
      \subcaption{Deeplearning - \#iter vs error rate\label{fig:dl-iters}}
    \end{boxedminipage}%%
    &%%
    \begin{boxedminipage}[b]{0.25\linewidth}%%
      \centering
      \includegraphics[width=0.9\linewidth]{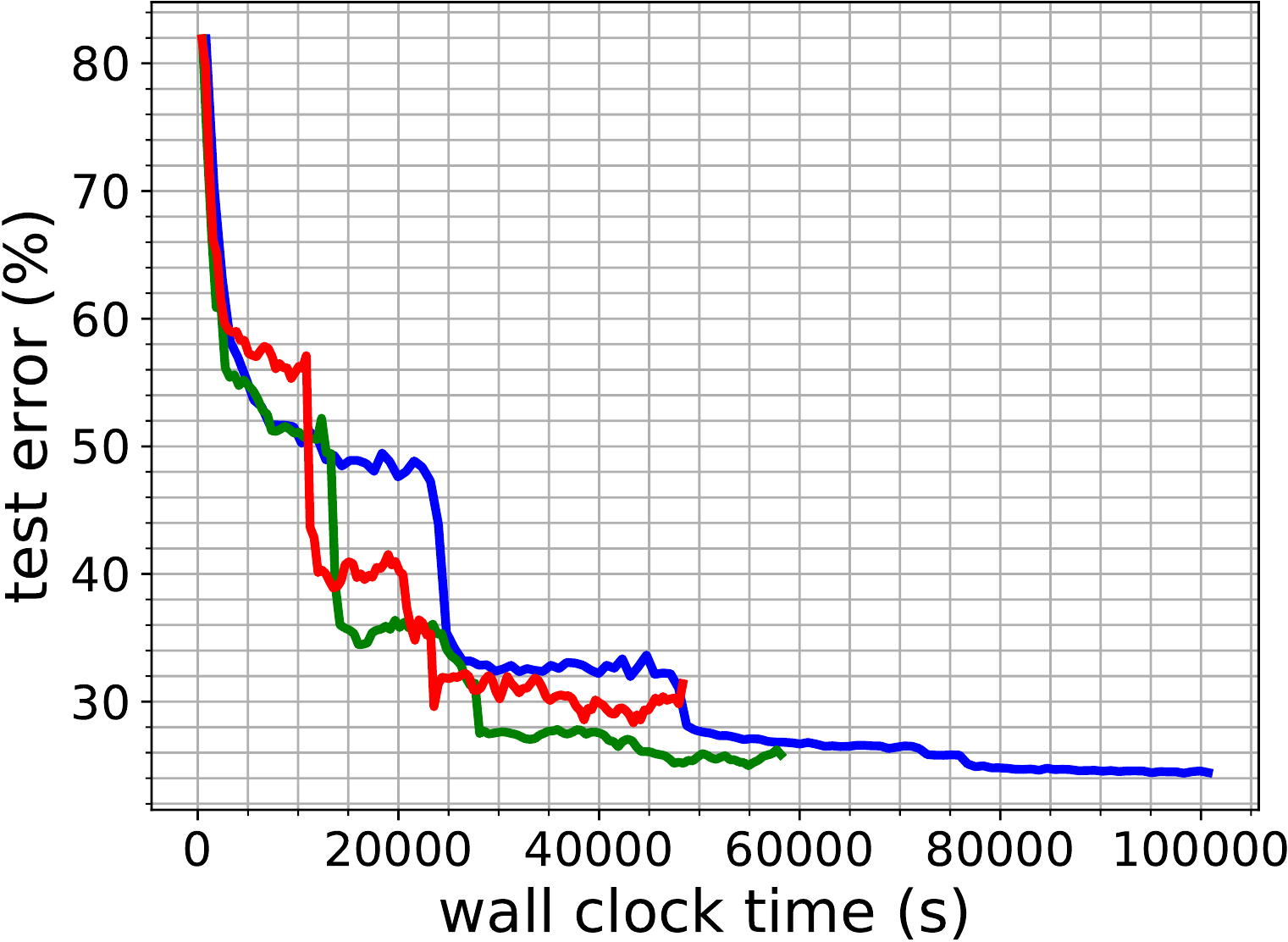}
      \subcaption{Deeplearning - time vs error rate\label{fig:dl-time}}
    \end{boxedminipage}%%
    &%%
    \begin{boxedminipage}[b]{0.25\linewidth}
      \centering
      \includegraphics[width=0.9\linewidth]{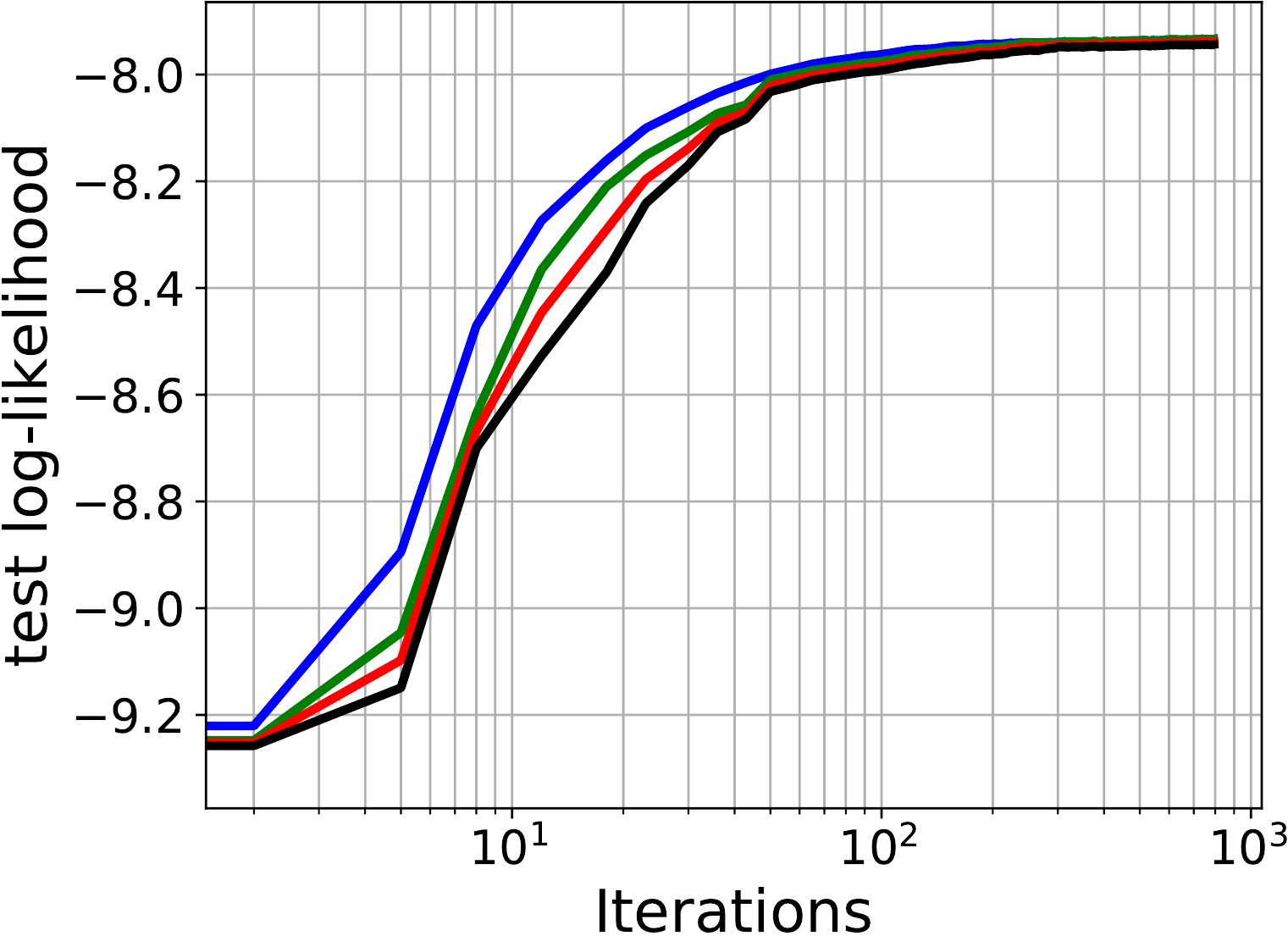}
      \subcaption{LDA - \#iter vs log-likelihood \label{fig:lda-iters}}
    \end{boxedminipage}%%
    &%%
    \begin{boxedminipage}[b]{0.25\linewidth}%%
      \centering
      \includegraphics[width=0.9\linewidth]{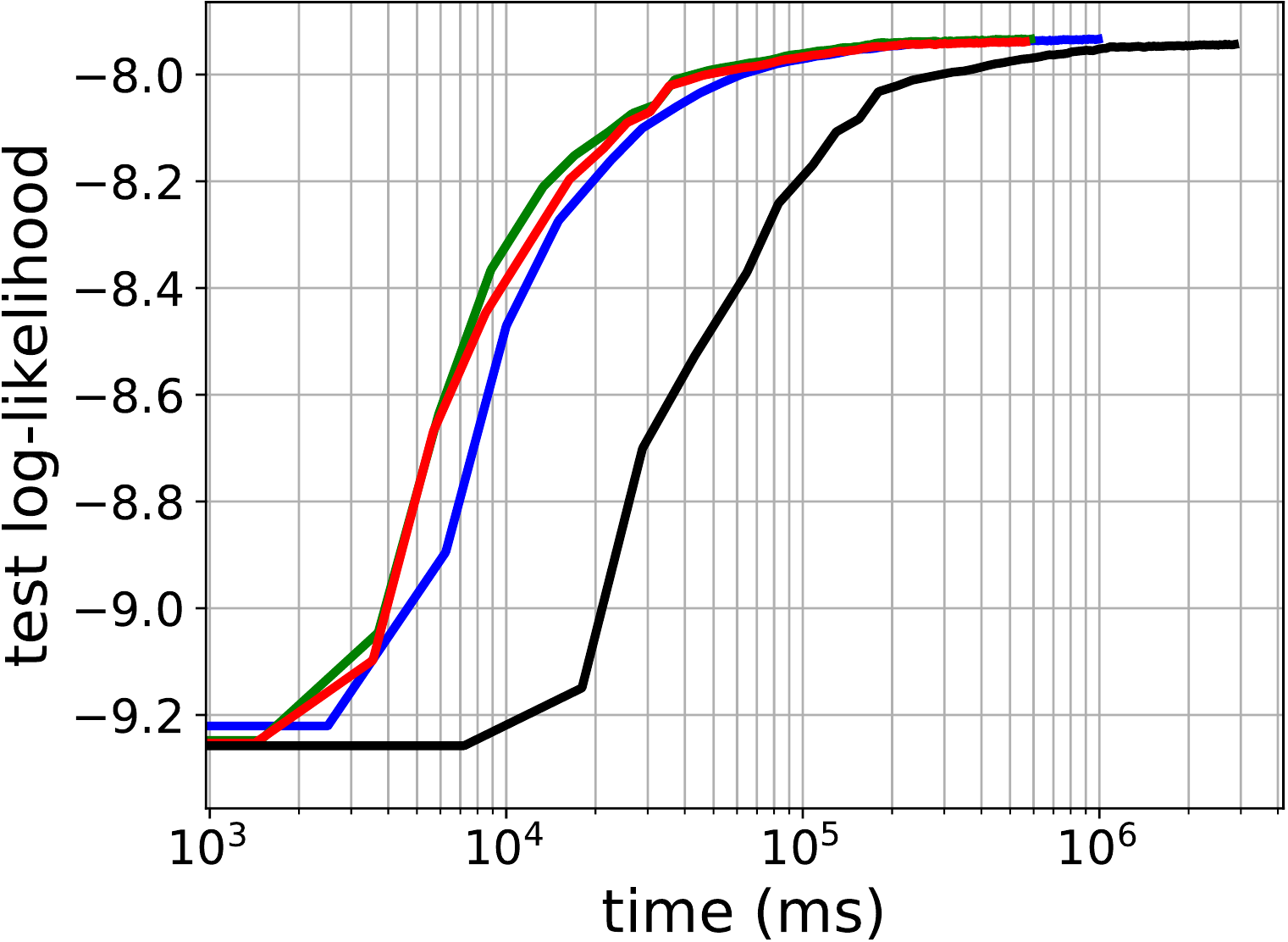}
      \subcaption{LDA - time vs log-likelihood \label{fig:lda-time}}
    \end{boxedminipage}%%
  \end{tabular}
%%   \vspace{-1ex}
  \caption{\sysname{} vs state-of-the-art approaches for asynchronous and synchronous LDA and Deep learning}
  \vspace{-0.1in}
\end{figure*}

\noindent
\textbf{Implementation:} \sysname{} is implemented in C++ as a thin
communication control layer between DML applications (e.g.,
PLDA~\cite{plda}, Keras~\cite{keras},
Tensorflow~\cite{tensorflow-osdi}) and MPI communication libraries
(OpenMPI~\cite{openmpi} and NCCL~\cite{nccl}). DML applications
interact with \sysname{} through APIs defined in
Table~\ref{tab:mlfabric-apis} and \sysname{} internally uses APIs
provided by MPI frameworks to aggregate/schedule transfers across
the network.

\noindent
\textbf{Datasets and ML models:} We evaluate \sysname{} with two
popular communication intensive distributed ML applications: (1)
distributed deep learning for image recognition on the
ImageNet1K~\cite{imagenet-data} data
set (using ResNet50 and ResNet152~\cite{resnet} models),
and, (2) distributed LDA for topic modelling
using Gibbs sampling on the NY Times dataset~\cite{nytimes}.
The computation and communication structure of
distributed LDA is similar to SGD; instead of computing a gradient
update using a mini-batch, each worker computes a numerical update to
a word-topic matrix using its entire shard of data, and then exchanges
the update among all the workers~\cite{plda} using a PS or MPI based
system.

\noindent \textbf{Experiment setup:} We run the DML applications
across 30 workers in a cluster of 15 baremetal machines connected by a
10 Gbps network.  The worker computation for distributed LDA runs on a
4-core CPU with 2.3GHz processor whereas the deep learning
applications use NVIDIA P100 GPUs (2 cards/physical machine).

The \coordinator{}, server (single) and replica (single) are hosted on a dedicated
machine with $10$Gbps bandwidth.  The aggregators are
co-hosted with worker clients; the aggregator runs on a separate core
on the worker machines and does not compete for CPU resources.
We batch requests at the \coordinator{} every $100 ms$.

\noindent
\textbf{Background compute and network load:} Along the lines of prior
work~\cite{ce-paper},
we emulate {\em compute stragglers} by slowing down the update
computation stage; a single worker has an $r$\% chance of being slowed
down by a factor of $s$; by default, $(r, s) = (10, 2)$ (compute
setting C1).  We also study other settings; C2: $(r,s) = (10, 4)$ and
C3: $(r,s)=(4,2)$.

We emulate {\em network background traffic} by varying the rate limits
on physical hosts' NICs; incoming and outgoing links are treated
independently.  For every $T$ (= 5, default) seconds the NIC rate is
changed to a value from the set $\{1, 2.5, 3.3, 5, 10\}$ Gbps with
probability $p=\{p_1, p_{2.5}, p_{3.3}, p_{5}, p_{10}\}$; it emulates
the case where there are $\{9, 3, 2, 1, 0\}$ other contending flows
respectively ($p = \{0, 0, 0, 0.1, 0.9\}$ is the default settings,
called N1). We also consider two other settings; N2:$(0, 0.1, 0.1,
0.1, 0.7)$ and N3:$(0.5,0,0,0,0.5)$.

The network monitor reports changes in link bandwidth to the \coordinator{}
after $t_{lag}$ (= 0.2s, default).

\noindent
\textbf{Algorithms:} We evaluate the following: PS-based asynchronous
and synchronous variants of \sysname{}, or {\sysname{}-A} \& {
  \sysname{}-S}, respectively; vanilla PS-based asynchronous ({\em Async}); and
MPI-based (using NCCL library) synchronous algorithms -- we study two
variants, ring-reduce and tree-reduce, or \emph{RR-Sync} and \emph{Tr-Sync}, respectively.

\subsection{Performance of \sysname{}-A}
\label{sec:eval-overall-benefits}

We compare \sysname{}-A with $Async$ and \emph{RR-Sync}.  We also study the effect of varying delay.

\noindent
\textbf{Distributed deep learning:}
Figures~\ref{fig:dl-iters}-~\ref{fig:dl-time} plot the top-1 test
error (in \%) for a ResNet50 model (100 MB) trained on the ImageNet1K
dataset as a function of training epochs and time respectively (for the setting
C1-N1).  We use a mini-batch size of 32 per worker and a learning
rate schedule that reduces by a factor of 10 after epochs 30, 60, and
90.  We compare \sysname{}-A only with \emph{RR-Sync}; the communication
bottleneck at the parameter server for $Async$ is prohibitive to run
even over a few days.

With a delay bound of 30, \sysname{}-A-30, can train a deep neural
network with 74\% accuracy; it alleviates server bottleneck
through update aggregation. The convergence rate as function of number
of epochs is similar to \emph{RR-Sync}.

In terms of wall clock time \sysname{}-A-30 is $1.74\times$ faster
than \emph{RR-Sync}.  The speed up can be attributed to: (1)
asynchronous algorithms are not prone to compute stragglers, (2)
unlike synchronous algorithms, \sysname{}-A does not have to send
traffic over low bandwidth links in each iteration; update from
workers behind slow links can be {\em dropped}. During the entire
training process, \sysname{}-A-30 dropped 30\% of the updates
at the worker for violating delay bounds.

We comment on the impact of delay control.  A higher delay bound than
the 30 used above can reduce the number of dropped updates and speed
up the training time at the cost of loss in accuracy.
\sysname{}-A-60, with a delay bound of 60, achieves only 70\% test
accuracy; however, it is $1.36\times$ times faster than
\emph{RR-Sync}. We also experimented with an intermediate delay (45)
and saw that it's accuracy and run-time lie between delay-30 and
delay-60.

\begin{table}
  \centering
  \scriptsize
      \begin{tabular}{c|c|c|c}
        & \textbf{NS1} & \textbf{NS2} & \textbf{NS3} \\ \hline
        \textbf{CS1} & 1.74 & 1.23 & 1.42 \\ \hline
        \textbf{CS2} & 2.96 & 2.0 &  2.32\\ \hline
        \textbf{CS3} & 1.90& 1.33& 1.42\\
      \end{tabular}
      \caption{Speedup in time w.r.t \emph{RR-Sync} \label{tab:speedup-stragglers}}
      % \vspace{-0.25in}
\end{table}

\noindent
\textbf{Varying compute and network settings:}
Next, we varied CS and NS  settings to
evaluate the benefits of \sysname{}-A for different kinds of heterogenuous
environments.
Table~\ref{tab:speedup-stragglers} shows the speedup of \sysname{}-A relative to {\em RR-Sync} %% async-\sysname{}
%% synchronous SGD
across 9 different compute and network background loads.  Here, to minimize
the overall running time, we start with a pre-trained model (i.e.,
the model after epoch 50 for synchronous SGD). The run time is measured as
time taken to reach 74\% test accuracy.

Speed-up is highest (3X) when some workers are $4\times$ slower than others (C2) and
the network is not the bottleneck (N1). This is because in \emph{RR-Sync}
AllReduce is triggered only after receiving update-ready notification from all
workers. Thus, in the presence of stragglers, network bandwidth remains fallow waiting for a slow worker to compute the update.

\noindent
\textbf{Varying reporting lag:} For network settings N2, increasing the
reporting lag up to 2s (from 0.2s, with the network re-configured every 5
seconds) increases the per-iteration time averaged over 10 epochs (for ResNet50) by 100ms--7.6\% of overall 1300ms.  However, for a skewed distribution
of link bandwidths, $p_{1}=0.2$ and $p_{10} = 0.8$, the per-iteration time
increased by 40\% with a 2s lag. Thus, the gains from \sysname{} are robust to
lags in monitoring unless bandwidth skews are significant.

\noindent
\textbf{Distributed LDA:} Figures~\ref{fig:lda-iters} and~\ref{fig:lda-time}
compares performance of \emph{RR-Sync}, \emph{Async} and \sysname{}-A based on
the number of iterations and time taken to converge for the topic modelling
task using NY times dataset (vocabulary size=102660, number of documents=300K);
we use compute setting C1 and network setting N1 for this experiment.  We learn
a model with 100 topics; the model is said to have converged when the
log-likelihood reaches -7.94 on a test data of size 1500.  \emph{RR-Sync},
\sysname{}-A-30, \sysname{}-A-60 and \emph{Async} converge in 145 (182s), 188
(139s), 239 (169s) and
300 (1080s) iterations (wall clock time) respectively. This corresponds to a
    $1.6\times$ and $1.25\times$ speedup (in number of iterations) for
    \sysname{} in comparison to \emph{Async} for with delay bounds 30 and 60
    respectively.  Further, even though \sysname{}-A takes more number of
    iterations, it reduces the overall run time w.r.t \emph{RR-Sync} by 24\% and 7\% for delays 30
    and 60. Due to update aggregation, \sysname is up to $7\times$ faster than
    \emph{Async}.
Similar gains were obtained for other compute and network settings.

\noindent
\textbf{Importance of delay control and aggregation:} Our results
above show the relative importance of these two aspects of
\sysname{}. They show that \sysname{}'s aggregation plays a crucial
role in supporting all large model training; without it, training is
prohibitively slow. Note that \sysname{} enables aggregation for the
first time for asynchronous algorithms. Delay control is also
important, because without it either accuracy (ResNet-50) or runtime
(LDA) are impacted.

\subsection{Performance of \sysname{}-S}
\label{sec:eval-aggregation}

We compare the performance of \sysname{}-S (using ResNet50) with
\emph{RR-Sync} for different compute and network settings.  We measure
the overall time to complete 5 epochs for both the algorithms.  We
find that the bandwidth optimal \emph{RR-Sync} is faster ($1.2-2
\times$) than \sysname{}-S for all combinations of compute and network
settings except C2 and N1.  When some workers are slowed down by a
factor of $4$, then for the C2-N1 setting, the rest of the workers are
idle (no computation or communication) for 50\% of the overall
runtime. For all other settings, communication generally is not
idle. \sysname{}-S reduces the idle time by eagerly aggregating
available worker updates (even over low bandwidth links).  This
results in a 16\% improvement in overall run time in C2-N1.  For
ResNet-152 model (240MB) with the above compute and network settings,
{\em RR-Sync} is the optimal algorithm since communication is always the
bottleneck.

\begin{figure}
\begin{tabular}{@{}c@{}c@{}}%%
  \begin{boxedminipage}[t]{0.49\linewidth}%%
    \centering%%
    \includegraphics[width=0.9\linewidth]{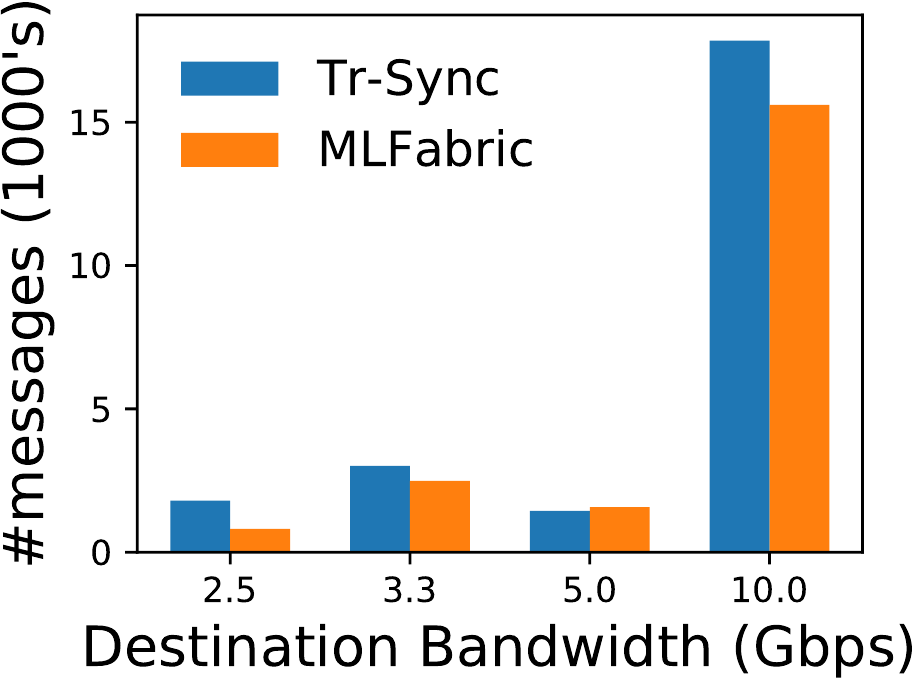}%%
    \caption{Histogram of number of update messages sent over links with different bandwidths. \label{fig:aggregation-histogram}}%%
  \end{boxedminipage}%%
  &%%
  \hspace{0.02\linewidth}%%
  \vspace{-1ex}
  \begin{boxedminipage}[t]{0.49\linewidth}%%
    \centering
    \includegraphics[width=0.9\linewidth]{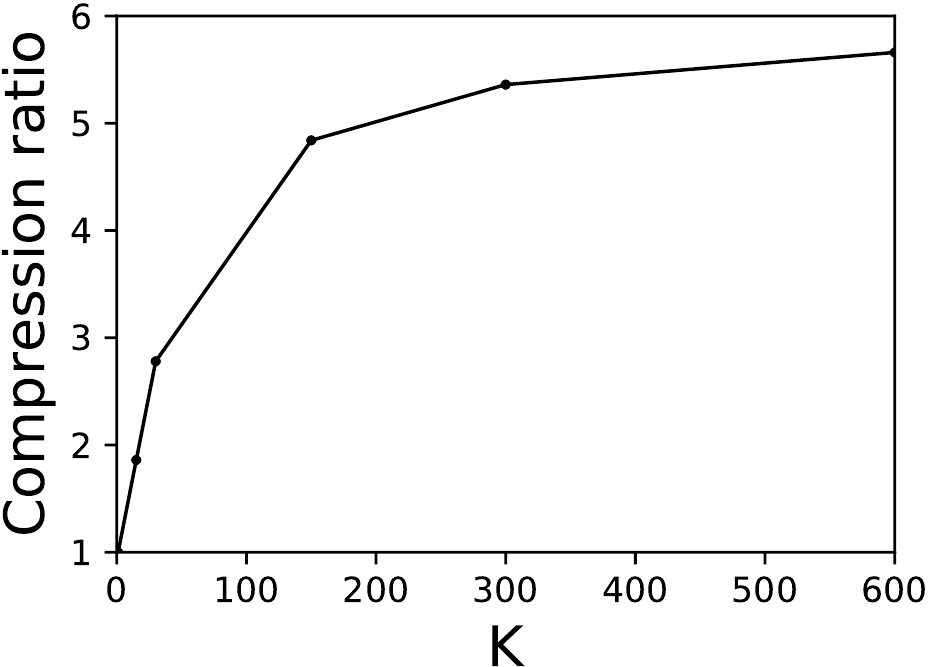}%%
    \caption{Reduction in data sent to replica as function of $Div_{max}$ (K) \label{fig:replication}}%$%
  \end{boxedminipage}%%
\end{tabular}%%
\end{figure}

We also compare \sysname{}-S with another (non-bandwith optimal) variant of MPI
AllReduce (\treereduce) that aggregates and distributes updates along a binary
tree.
In the presence of stragglers (C2) and network contention (N2),
\sysname{}-S reduces the per-iteration communication time for
ResNet-152 by 21.7\%: from 3.05s with \treereduce to 2.38s. For
ResNet-50 the gain is 18.42\%.  Clearly, the advantages of
network-aware aggregation are more prominent for larger model sizes.
Since compute time is relatively small, the reduction in per-iteration
time directly translates to reduction in overall running time of the
algorithm for large models.  The benefits arise from dynamically
avoiding (aggregating more) updates at nodes with low (high) current
bandwidth; figure~\ref{fig:aggregation-histogram} plots the number of
updates aggregated as a function of the bandwidth available on the
incoming link.  Note that since \treereduce uses a binary tree the
number of messages exchanged between workers is higher. However, being
network-aware, \sysname{}-S forwards only 816 of the overall 20000 (3\%)
messages to aggregators with low bandwidth (2.5Gbps in N2), whereas
\treereduce sends 1800 messages over such links!

\subsection{Bounded consistency replication}
\label{sec:eval-replication}

For MPI based systems, fault tolerance is provided today by
checkpointing the model at the worker with rank 0. We compare the cost
of checkpointing with the cost of transferring updates to a
hot-standby replica over the network. We measure the overhead of
fault-tolerance as follows: for MPI based systems the overhead is the
difference in time between two runs with and without checkpointing
over 6 epochs ($\approx 20$ minutes with no stragglers/network
bottlenecks). For PS based systems, it is the time
difference between two runs with and without in-network replication.
The runs with fault tolerance are parameterized by the maximum
allowable divergence (measured in number of updates) between server and
replica; for MPI systems it translates to checkpointing frequency.
The overhead of checkpointing for every iteration and every 20
iterations is 76 minutes and 4 minutes, respectively, for MPI based
systems; the corresponding overhead for in-network replication is 16
mins and 10 mins, respectively. As the divergence bound increases (600
updates in the case of 20 iterations), the network cost savings due to
aggregation plateaus at a factor of $5.6$ for 30 workers (see
fig.~\ref{fig:replication}). Further, if all workers only write part of
their model to disk every iteration and replicate it 3-way, the 76
minute overhead can be reduced to just $7$ minutes. Thus, overall,
in-network replication does {\em not} help with MPI based systems.

For PS based DML systems running asynchronous SGD, checkpointing at
the server for every 30 and 600 updates has an overhead of 96 minutes
(6X worse w.r.t. in-network replication) and 6 minutes (0.6X),
respectively. Thus, in-network replication is advantageous over
checkpointing here for scenarios that warrant tight divergence bound
(e.g., where compute nodes are highly susceptible to
failure).

Chain replication is commonly used in many PS frameworks. We also
experimented with it, but found that, given the large model sizes, it
adds prohibitive overhead (up to 30X) compared to in-network
replication in \sysname{}.

\noindent
\subsection{\Coordinator{} performance}
\label{sec:coordinator-perf}
For our experimental setting with 30 workers, the transfer schedules were
computed within $5$~ms per batch (batch size, $|U|$, is typically $<10$).  To
test scalability of the \coordinator{} computation\footnote{Because the \coordinator{} processes only small control
messages that are received on a dedicated TCP socket and take only 1
RTT, high network utilization does not affect \coordinator{} response
time.}, we studied the effect of
varying sizes of $|U|$. We  measure the time taken by the
\coordinator{} to determine the concrete batch schedules by providing batches of
$|U|$ updates (with random deadlines from $Uniform(1, 2|U|)$) for a network
topology with $|U|/2$ nodes and a congestion free core.
For update batch sizes of 100,
500 and 1000, the \coordinator{} overhead was $30$~ms, $440$~ms, and $1460$~ms,
    respectively. Thus, the  overhead is quadratic \textit{w.r.t.}
    update batch size. However, we note that the inner loops in
    alg.~\ref{code:final-ordering-algorithm} (line 3, function \texttt{ShrtUp})
    and~\ref{code:aggregation-algorithm} (lines 21-23) can be parallelized (which
    our current implementation does not),  leading to better scaling.

% !TEX root = main.tex
\section{Related work}
\label{sec:related}

Prior works propose various techniques to reduce the overall training
time of ML algorithms that employ SGD for learning.

\noindent
{\bf Algorithmic approaches:} Some other approaches for mitigating
stragglers involve: aggregating gradients from only a subset of fast
workers in each iteration of synchronous SGD~\cite{distbelief}, which
is complementary with \name's aggregation; and \emph{delay}-aware
learning rate for asynchronous SGD~\cite{adadelay}, which can benefit
from \name's delay management. Prior work advocates \emph{variance
  reduction} SGD where a series of asynchronous updates is
interspersed with intermediate synchronous updates~\cite{distbelief},
and performing partial updates of the model to reduce total data sent
over the network~\cite{bosen,strads}; both techniques can benefit from
\name.

\noindent
{\bf System-level approaches:} To speed up gradient computation time,
ML systems leverage SIMD processing capabilities of hardware
accelerators like GPU or TPU~\cite{tensorflow-osdi}, which can
leverage \sysname{} for further speedup.  Communication overhead is
typically managed by: (1) workers leveraging sparsity of data and
\emph{pulling} only parts of the model \cite{parameterserver-osdi}, or
(2) quantization of floating point values used to represent gradients
\cite{one-bit-sgd}.
\name is complementary with \#1 and \#2.

Our overall approach can be view as type of
co-flow~\cite{coflow,varys,aalo,coflow-approx,rapier} scheduling.  The
differences in our setting are: (1) flows in our co-flow have an
intrinsic order, (2) we can drop/re-order flows from the co-flow, (3)
flows in our co-flow can be aggregated in-network using ML algorithm
specific aggregation functions. These aspects make our problem
markedly difficult and different.

\section{Conclusion}
\label{sec:conclusion}

 We designed \sysname{}, a communication library for speeding up
 large-scale distributed machine learning (DML) systems in dynamic
 cluster settings. We showed that fine-grained in-network control
 helps \sysname{} to (1) algorithmically speed up convergence, (2)
 improve network efficiency via dynamic update aggregation, and (3)
 offload model replication responsibilities from servers to the
 network in a network-efficient manner. Our experiments on a 30-worker
 GPU cluster using real-world datasets and realistic straggler settings
 show that \sysname{} reduces model training time by up to $3\times$
 compared to state-of-the-art algorithms. Finally, this work does not raise any
 ethical issues.

% \newpage
\section{Appendix}

% !TEX root = main.tex

\subsection{ILP formulation for joint ordering and forwarding for aggregation}
\label{sec:one-shot-optimization}

Let $W = \{w_1,..,w_n\}$ be the workers and $S$ be the server storing
a DML application's model.  Let, $A =\{a_1,..,a_{\ell}\}$ be the
aggregators that serve as intermediate hops.  Let $G= (V,E)$ denote a
directed graph representing the underlying communication network. $V$
is the set of all hosts and switches and $E$ is the set of network
links including host to switch links. Let $B(e), \forall e \in E$
denote the capacity of link $e$.
We assume the path,  $P(v_1, v_2)$, from $v_1$ to $v_2$ over the set of links
$E$ is fixed and does not change in time. Different paths can share a
network link.

To exploit dynamic aggregation and re-ordering, we \emph{jointly} determine
the schedule for a batch of requests\footnote{requests are batched temporally,
so that earlier requests are not starved}; let $U = \{g_1, \ldots, g_{|U|}\}$
denote a batch of ready updates.

\noindent
\textbf{Variables:} Let the variable $r_{g_i}(t)$ denote the rate at
which update $g_i$ is transmitted by $w_i$ over time. Modeling the
rate as a function of time allows us to capture network time sharing
and ordering between updates. E.g., if updates $g_1$ and $g_2$ from
$w_1$ and $w_2$ have to time-share a link $e$ such that $w_1$
sends data first followed by $w_2$, the rate for $g_1$ and $g_2$ are:
\begin{equation}
r_{g_1}(t) =
  \begin{cases}
    B(e), & \text{if}\quad 0 \leq t < \frac{sz(g_1)}{B(e)} \\
    0, & \text{otherwise}
  \end{cases}
\end{equation}
\begin{equation}
r_{g_2}(t) =
  \begin{cases}
    B(e), & \text{if}\quad \frac{s(g_1)}{B(e)} \leq t < \frac{sz(g_1)
      + sz(g_2)}{B(e)} \\ 0, & \text{otherwise}
  \end{cases}
\end{equation}
Here, $sz(g)$ is the update size. Update ordering hinges on start/end
times:
\begin{eqnarray}
t_{st}(g_i) & = & \{ \text{min}(t) : r_{g_i}(t) > 0\} \\
t_{en}(g_i) & = & \{ \text{max}(t) : r_{g_i}(t) > 0\}
\end{eqnarray}
$g_1$ is applied before $g_2$ if $t_{en}(g_1) < t_{en}(g_2)$.

Let the integer variable $dst(g_i)$ denote the immediate next hop for $g_i$; since
a worker can forward the update either to the server or an aggregator,
$dst(g_i) \in \{S\} \cup A$.
We also determine the schedule of aggregated updates.
Let $r_{a_j}(t)$ be the rate at which $a_j$ forwards the aggregated update to
the server and let $t_{st}(a_j)$ and $t_{en}(a_j)$ represent the start/end
times.

\noindent
\textbf{Objective:}
For synchronous SGD, we have:
\begin{equation}
  \text{obj}_{sync} := \max \left(\max_{\substack{g \in U \\ dst(g) = s}} t_{en}(g),\quad \max_{a \in A}~~t_{en}(a) \right)
\end{equation}
This minimizes the total time to aggregate all updates in $U$.  For
asynchronous SGD, we optimize the average completion time per update:
\begin{equation}
  \text{obj}_{async} := \frac{1}{|U|} \left(  \summation{\substack{g \in U \\ dst(g) = s}}{}
    t_{en}(g) + \summation{a \in A}{} m(a) t_{en}(a) \right)
\end{equation}
where, $m(a)$ is the number of updates aggregated at $a$. Further, $t_{en}(g),
\forall g \in U$, should be such that delay bounds are satisfied.

Modeling the destination for each update ($dst(g)$), rate of transfer at each
discrete point in time ($r_g(t)$) results in an large number of discrete
variables. Solving an ILP with large number variables is time-wise expensive
and is not a straight forward choice for the low-latency requirements at the
\coordinator{}.
Thus, \name  breaks down the complex ILP into smaller sub-problems (ordering,
aggregation, replication) and develops computationally efficient heuristics to
solve them.
%% with determining the time-varying rate that minimizes completion time,
%% while ensuring ordering guarantees makes the problem
%% intractable.

\subsection{Model split across multiple servers}
\label{sec:method-multiple-servers}

Our algorithms in \secref{sec:method} considered the case of a single
parameter server; we now briefly describe the case where the model is
split across a set $S$ of servers.  Thus, $g_i$, from a worker
consists of $|S| > 1$ components, $g_i = \{g_i^1, \ldots,
g_i^{|S|}\}$.  All the components of $g_i$ are computed from the same
version of the model and thus, will have the same deadline. Below, we
ignore aggregation and replication and consider just scheduling
(\secref{sec:method-ordering-updates}) for simplicity; the
modifications we suggest below naturally apply to algorithms for
replication (\secref{sec:method-multiple-replicas}) and aggregation
(\secref{sec:in-network-aggregation}).

One option to schedule updates to multiple servers is to use an algorithm
similar to one described in \secref{sec:method-bounding-delay} by defining
deadlines for each individual server and update component.  For example,
consider two updates, $g_1 = [g_1^1, g_1^2]$ and $g_2 = [g_2^1, g_2^2]$, to two
destination servers, $s_1$ and $s_2$. By treating all the updates as
independent and choosing them in a shortest transfer first order we might
reserve network resources in the following order $g_1^1 \rightarrow g_2^1
\rightarrow g_1^2 \rightarrow g_2^2$, if updates to $s_1$ are small in size. In
the presence of a large number of workers, this would result in some parts of
the model being updated less frequently than others.

To ensure uniform number of updates to all components of the ML model,
network resources for all update components are reserved together. In
each iteration of our algorithm, we pick the update which has the
largest completion time for all its components:
\begin{equation}
t_{en}(g) = \max_{j}~ t_{en}(g^j)
\end{equation}

\subsection{Model distribution}
\label{sec:method-model-distribution}

Aggregating updates reduces overall
runtime by reducing the amount of data forwarded to the server.  However, if
each request for the model is responded to individually then the down-link at
the server will become the bottleneck. To reduce the load on the down-link, we
use a distribution tree for propagating the model to the workers.  At the
server, requests are batched and responded with same version of model.
%% ; each group is returned with the same version of the model.
The distribution pattern is determined similar to the aggregation pattern. For
a batch of requests, $k^{\circ}$ \emph{distributors} are earmarked.  Mapping of
workers to distributors is done using a variant of
alg.~\ref{code:aggregation-algorithm} obtained by replacing $t_{en}()$s as the
times taken to transfer the model from server-to-distributor and
distributor-to-worker.  Once the partitioning is determined, we first transfer
the  model  from the server to the $k^{\circ}$-th distributor and then proceed
backwards.  The workers in the first group receive the model directly from the
server.

\newpage
\twocolumn[
\begin{@twocolumnfalse}
\subsection{SGD convergence analysis under bounded delay}
\label{sec:proof-convergence}
{
  \footnotesize
We extend the proof of convergence under uniform delay from \cite{adadelay}.
Specifically, we modify Lemma A.3, under the assumption that delay is uniform
in: $\tau_{t,\epsilon} \sim \textrm{Uniform} ({\bar{\tau}-\epsilon,
\bar{\tau}+\epsilon})$. We bound the delay dependent term $\Delta(t, \epsilon)$
(see A.15 in \cite{adadelay}, also defined below) under the new delay model.

\begin{equation}
\Delta(t,\epsilon) := \frac{1}{2\alpha\big(t,\tau(t,\epsilon)\big)}\big[||x^*-x_t||^2 -||x^*-x_{t+1}||^2\big]
\end{equation}

We then show that,

\begin{equation}
\sum_{t=1}^T\mathbb{E}[\Delta(t,\epsilon)] \leq \frac{1}{2}(L+c)R^2+\mathcal{O}(\epsilon\sqrt{T-\epsilon})
\end{equation}

The expected loss after $T$ iterations can then be bounded as (see Corollary 3.2
in \cite{adadelay}):

\begin{equation*}
  E[L(\wb_T)] - L(\wb^*) \leq \frac{1}{t}
  \sum_{t=1}^T\mathbb{E}[\Delta(t,\epsilon)] \leq \frac{1}{T}
  \mathcal{O}(\epsilon\sqrt{T-\epsilon})\quad\quad\hfill
\end{equation*}

\textit{Proof}:
Let $r_t=||x_t-x^*||^2$, observe that it is not independent of $\tau(t-1,\epsilon)$. Thus, with
\begin{equation}
z_{t,\epsilon}=\frac{1}{\alpha\big(t,\tau(t,\epsilon)\big)} - \frac{1}{\alpha\big(t-1,\tau(t-1,\epsilon)\big)} = c\Big(\sqrt{t+\tau(t,\epsilon)}-\sqrt{t-1+\tau(t-1,\epsilon)}\Big)
\end{equation}
we have
\begin{equation}
\sum_{t=1}^T\mathbb{E}[\Delta(t,\epsilon)] = \mathbb{E}\bigg[\sum_{t=1}^T\Delta(t,\epsilon)\bigg]
=\frac{1}{2}\mathbb{E}\bigg[\frac{r(1)}{\alpha\big(1,\tau(1,\epsilon)\big)}+\sum_{t=1}^T z_{t,\epsilon}r_t\bigg]
\leq \frac{1}{2}(L+c)R^2+\frac{1}{2}\mathbb{E}\bigg[\sum_{t=2}^T z_{t,\epsilon}r_t\bigg]
\end{equation}

\begin{equation}
\mathbb{E}[z_{t,\epsilon}r_t] = \mathbb{E}_{\tau_{t,\epsilon}}\big[\mathbb{E}[z_{t,\epsilon}r_t|\tau_{t,\epsilon}]\big] =  \frac{1}{2\epsilon+1}\sum_{l = \bar{\tau}-\epsilon}^{\bar{\tau}+\epsilon}\bigg(\sum_{s=\bar{\tau}-\epsilon}^{l-1}\frac{r_{t,s}c}{2\epsilon+1}\Big(\sqrt{t+l}-\sqrt{t-1+s}\Big)\bigg)
\end{equation}

Consider the inner summation, we have
\begin{equation}
\frac{cR^2\sum_{s={\bar{\tau}-\epsilon}}^{l-1}(l-s+1)}{(2\epsilon+1)\sqrt{2(t+\bar{\tau}-\epsilon)-1}} = \frac{cR^2(\bar{\tau}-\epsilon+l)(l+3+\bar{\tau}-\epsilon)}{2\cdot(2\epsilon+1)\sqrt{2(t+\bar{\tau}-\epsilon)-1}}
\end{equation}

Thus, we now consider
\begin{align*}
\mathbb{E}[z_tr_t] &\leq \frac{1}{2\epsilon+1}\sum_{l = \bar{\tau}-\epsilon}^{\bar{\tau}+\epsilon}\frac{cR^2(\bar{\tau}-\epsilon+l)(l+3+\bar{\tau}-\epsilon)}{2\cdot(2\epsilon+1)\sqrt{2(t+\bar{\tau}-\epsilon)-1}}\\
&=\frac{cR^2}{2\cdot (2\epsilon+1)^2\sqrt{2(t+\bar{\tau}-\epsilon)-1}}\sum_{l = \bar{\tau}-\epsilon}^{\bar{\tau}+\epsilon}(\bar{\tau}-\epsilon+l)(l+3+\bar{\tau}-\epsilon)
\end{align*}

Since
\begin{equation}
\sum_{l = \bar{\tau}-\epsilon}^{\bar{\tau}+\epsilon}(\bar{\tau}-\epsilon+l)(l+3+\bar{\tau}-\epsilon) = \frac{2}{3}(2\epsilon+1)(9\bar{\tau}+6\bar{\tau}^2-4\epsilon-6\bar{\tau}\epsilon+2\epsilon^2)
\end{equation}

\begin{align*}
\mathbb{E}[z_tr_t] &=\frac{cR^2(9\bar{\tau}+6\bar{\tau}^2-4\epsilon-6\bar{\tau}\epsilon+2\epsilon^2)}{3\cdot (2\epsilon+1)\sqrt{2(t+\bar{\tau}-\epsilon)-1}}
<\frac{cR^2(9\bar{\tau}+6\bar{\tau}^2)}{3\cdot (2\epsilon+1)\sqrt{2(t+\bar{\tau}-\epsilon)-1}}
+ \frac{cR^2(-4\epsilon-6\bar{\tau}\epsilon+2\epsilon^2)}{3\cdot 2\epsilon\sqrt{2(t+\bar{\tau}-\epsilon)-1}}\\
&=\frac{cR^2(9\bar{\tau}+6\bar{\tau}^2)}{3\cdot (2\epsilon+1)\sqrt{2(t+\bar{\tau}-\epsilon)-1}}
-\frac{cR^2(4\epsilon+6\bar{\tau}\epsilon)}{3\cdot 2\epsilon\sqrt{2(t+\bar{\tau}-\epsilon)-1}}
+\frac{cR^22\epsilon^2}{3\cdot 2\epsilon\sqrt{2(t+\bar{\tau}-\epsilon)-1}}\\
&<\frac{cR^2(9\bar{\tau}+6\bar{\tau}^2)}{3\cdot 1\sqrt{2(t+\bar{\tau}-\epsilon)-1}}
-\frac{cR^2(2+3\bar{\tau})}{3\cdot \sqrt{2(t+\bar{\tau}-\epsilon)-1}}
+\frac{cR^2\epsilon}{3\cdot \sqrt{2(t+\bar{\tau}-\epsilon)-1}}\\
&=\frac{cR^2(6\bar{\tau}+6\bar{\tau}^2-2)}{3\sqrt{2(t+\bar{\tau}-\epsilon)-1}}
+\frac{cR^2\epsilon}{3\sqrt{2(t+\bar{\tau}-\epsilon)-1}}
\end{align*}

Finally,
\begin{equation}
\mathbb{E}\bigg[\sum_{t=2}^T z_{t,\epsilon}r_t\bigg] = \sum_{t=2}^T\mathbb{E}[z_{t,\epsilon}r_t]
\leq \int_1^T\frac{cR^2(6\bar{\tau}+6\bar{\tau}^2-2)}{3\sqrt{2(t+\bar{\tau}-\epsilon)-1}}
+\frac{cR^2\epsilon}{3\sqrt{2(t+\bar{\tau}-\epsilon)-1}}\mathrm{d}t = \mathcal{O}(\epsilon\sqrt{T+\bar{\tau}-\epsilon})
\end{equation}

}

\end{@twocolumnfalse}
]
\newpage
\begin{footnotesize}
  \bibliographystyle{acm}
  \bibliography{distributed_ml}
\end{footnotesize}
\end{document}